\documentclass[11pt,a4paper,final]{article}
\usepackage{amsmath}%
\usepackage{amsthm}%
\usepackage{amstext}%
\usepackage{amssymb}%
\usepackage{showkeys}%
\usepackage{epsfig}%
\usepackage{graphicx}%
\usepackage{multicol}
\usepackage{xcolor}
\usepackage{apacite}
\usepackage[top=1in, bottom=1.25in, left=0.8in, right=0.8in]{geometry}

\newtheorem{thm}{Theorem}
\newtheorem{cor}{Corollary}[section]
\newtheorem{lem}{Lemma}
\newtheorem{rem}{Remark}

\begin{document}

\title{Options Pricing under Bayesian MS--VAR Process}
\author{Battulga Gankhuu\footnote{
Department of Applied Mathematics, National University of Mongolia, Ulaanbaatar, Mongolia;
E-mail: battulga.g@seas.num.edu.mn}}
\date{}

\maketitle 

\begin{abstract}
In this paper, we have studied options that are based on a Bayesian Markov--Switching Vector Autoregressive (MS--BVAR) process using a risk--neutral valuation approach. A BVAR process, which is a special case of the Bayesian MS--VAR process is widely used to model interdependencies of economic variables and forecast economic variables. Here we assumed that a regime--switching process is generated by a homogeneous Markov process and a residual process follows a conditional heteroscedastic model. With a direct calculation and change of probability measure, for some frequently used options, we derived pricing formulas. An advantage of our model is it depends on economic variables and is easy to use as compared to previous option pricing papers, which depend on regime--switching. \\[3ex]

\end{abstract}

\section{Introduction}

The first option pricing formula dates back to classic papers of \citeA{Black73} and \citeA{Merton73}. They implicitly introduced a risk--neutral valuation method to arbitrage pricing. But it was not fully developed and appreciated until the works of \citeA{Harrison79} and \citeA{Harrison81}. The basic idea of the risk--neutral valuation method is that discounted price process of an underlying asset is a martingale under some risk--neutral probability measure. The option price is equal to an expected value, with respect to the risk--neutral probability measure, of discounted option payoff. In this paper, to price derivative, we use the risk--neutral valuation method in the presence of economic variables. 

Sudden and dramatic changes in the financial market and economy are caused by events such as wars, market panics, or significant changes in government policies. To model those events, some authors used regime--switching models. The regime--switching model was introduced by seminal works of \citeA{Hamilton89,Hamilton90} (see also books of \citeA{Hamilton94} and \citeA{Krolzig97}) and the model is hidden Markov model with dependencies, see \citeA{Zucchini16}. However, Markov regime--switching models have been introduced before Hamilton (1989), see, \citeA{Goldfeld73}, \citeA{Quandt58}, and \citeA{Tong83}. The regime--switching model assumes that a discrete unobservable Markov process generates switches among a finite set of regimes randomly and that each regime is defined by a particular parameter set. The model is a good fit for some financial data and has become popular in financial modeling including equity options, bond prices,  and others. 

According to \citeA{Hardy01}, for monthly TSE 300 and S\&P 500 index returns, there is evidence that a two--regime model provides a good fit. Recently, to model required rate of return on stock, \citeA{Battulga23b} applied a two--regime model. The result of the paper reveals that the regime--switching model is good fit for the required rate of return. \citeA{Bollen98} used the lattice method and simulation for pricing American and European options with two regime--switching. To price some exotic options, \citeA{Boyle07} used regime--switching for underlying asset volatility to obtain partial differential equations. \citeA{Guo01} used regime--switching to model European option with complete and inside information. To price European option, \citeA{Hardy01} developed a recursive approach based on total sojourn random variable in the regime--switching framework. \citeA{Yao06} used a successive approximating scheme to price European option in regime--switching with continuous time Markov chain. For bond pricing with regime--switching, see \citeA{Bansal02} and \citeA{Landen00}. For the regime--switching models, the drift and volatility parameters of an underlying asset depend on a Markov chain. However, when the parameters are modeled by the Markov chain, the valuation of the options becomes complex. 

Economic variables play important roles in any economic model. In some existing option pricing models, the underlying asset price is governed by some stochastic process and it has not taken into account economic variables such as GDP, inflation, unemployment rate, and so on. For example, the classical Black--Scholes option pricing model uses a geometric Brownian motion to capture underlying asset price. However, the underlying asset price modeled by geometric Brownian motion is not a realistic assumption when it comes to option pricing. In reality, for the Black--Scholes model, the price process of the asset should depend on some economic variables. 

Classic Vector Autoregressive (VAR) process was proposed by \citeA{Sims80} who criticize large--scale macroeconometric models, which are designed to model interdependencies of economic variables. Besides \citeA{Sims80}, there are some other important works on multiple time series modeling, see, e.g., \citeA{Tiao81}, where a class of vector autoregressive moving average models was studied. For the VAR process, a variable in the process is modeled by its past values and the past values of other variables in the process. After the work of \citeA{Sims80}, VARs have been used for macroeconomic forecasting and policy analysis. However, if the number of variables in the system increases or the time lag is chosen high, then too many parameters need to be estimated. This will reduce the degrees of freedom of the model and entails a risk of over--parametrization. 

Therefore, to reduce the number of parameters in a high--dimensional VAR process, \citeA{Litterman79} introduced probability distributions for coefficients that are centered at the desired restrictions but that have a small and nonzero variance. Those probability distributions are known as Minnesota prior in Bayesian VAR (BVAR) literature, which is widely used in practice. Due to over--parametrization, the generally accepted result is that forecast of the BVAR model is better than the VAR model estimated by the frequentist technique. The BVAR relies on Monte--Carlo simulation methods. Recently, for Bayesian Markov--Switching VAR process, \citeA{Battulga24g} introduced a new Monte--Carlo simulation method that removes duplication in a regime vector. Also, the author introduced importance sampling method to estimate probability of rare event, which corresponds to endogenous variables. Research works have shown that BVAR is an appropriate tool for modeling large data sets, for example, see \citeA{Banbura10}.

In this paper, to partially fill the gaps mentioned above, we introduced Bayesian Markov--Switching VAR (MS--BVAR) model to value derivatives. Our model offers the following advantages: (i) it tries to mitigate the valuation complexity of previous derivative pricing models with regime--switching (ii) it considers economic variables thus the model will be more consistent with future economic uncertainty (iii) it introduces regime--switching so that the model takes into account sudden and dramatic changes in the economy and financial market (iv) it adopts a Bayesian procedure to deal with over--parametrization. The novelty of the paper is that we introduced Bayesian MS--VAR process, which is widely used to model economic variables to derivative pricing. 
Our model talks about not only the Bayesian MS--VAR process but also conditional heteroscedastic models for a residual process. The ARCH model proposed in \citeA{Engle82} is the prototype of all conditional heteroscedastic models and is commonly used to model time--varying volatility and volatility clustering, which are stylized facts of financial time series, see \citeA{McNeil05}.

The rest of the paper is structured as follows. Section 2 provides the main theorems and corollary, which will be used for options pricing. In section 3, we consider the domestic market and obtain pricing formulas for arithmetically weighted Black--Scholes European options using the normal distribution. In section 4, we focus on a domestic--foreign market, where assets take positive values.  Here we study some probability measures, which is originated from a risk--neutral probability measure and obtain valuation formulas for general European options. Section 5 provides some term structure models. Here we obtain pricing formulas for caplet, floorlet, and zero--coupon bond options. Next, we provide a conclusion in section 6. Finally, we give proofs of the theorems, corollary, and some useful lemmas.

\section{Main Results}

Let $(\Omega,\mathcal{H}_T,\mathbb{P})$ be a complete probability space, where $\mathbb{P}$ is a given physical or real--world probability measure. Other elements of the probability space will be defined below. To introduce a regime--switching in option pricing, we assume that $\{s_t\}_{t=1}^T$ is a homogeneous Markov chain with $N$ state and $\mathsf{P}:=\{p_{ij}\}_{i=0,j=1}^N$ is a random transition probability matrix, including an initial probability vector, where $\{p_{0j}\}_{j=1}^N$ is the initial probability vector. We consider a Bayesian Markov--Switching Vector Autoregressive (MS--VAR($p$)) process of $p$ order, which is given by the following equation 
\begin{equation}\label{01001}
y_t=A_{0,s_t}\psi_t+A_{1,s_t}y_{t-1}+\dots +A_{p,s_t}y_{t-p}+\xi_t,~t=1,\dots,T,
\end{equation}
where $y_t=(y_{1,t},\dots,y_{n,t})'$ is an $(n\times 1)$ vector, $\psi_t=(1,\psi_{2,t},\dots,\psi_{k,t})'$ is a $(k\times 1)$ random vector of exogenous variables, $\xi_t=(\xi_{1,t},\dots,\xi_{n,t})'$ is an $(n\times 1)$ residual process, $A_{0,s_t}$ is an $(n\times k)$ random coefficient matrix at regime $s_t$ that corresponds to the vector of exogenous variables, for $i=1,\dots,p$, $A_{i,s_t}$ are random $(n\times n)$ coefficient matrices at regime $s_t$ that correspond to $y_{t-1},\dots,y_{t-p}$. It should be noted that in general, the order $p$ can be random but to reduce the computational burden we do not take into account this case. Equation \eqref{01001} can be compactly written by
\begin{equation}\label{01002}
y_t=\Pi_{s_t}\mathsf{Y}_{t-1}+\xi_t,~t=1,\dots,T,
\end{equation}
where $\Pi_{s_t}:=[A_{0,s_t}: A_{1,s_t}:\dots:A_{p,s_t}]$ is a random coefficient matrix at regime $s_t$, which consist of all the random coefficient matrices and $\mathsf{Y}_{t-1}:=(\psi_t',y_{t-1}',\dots,y_{t-p}')'$ is a vector, which consist of exogenous variable $\psi_t$ and last $p$ lagged values of the process $y_t$. In the paper, this form of the MS--BVAR process $y_t$ will play a major role than the form, which is given by equation \eqref{01001}.

For the residual process $\xi_t$, we assume that it has $\xi_t:=\Sigma_t^{1/2}\varepsilon_t \in\mathbb{R}^n$, $t=1,\dots,T$ representation, see \citeA{Lutkepohl05} and \citeA{McNeil05}, where $\Sigma_t^{1/2}$ is Cholesky factor of a positive definite random matrix $\Sigma_t\in\mathbb{R}^{n\times n}$, which is measurable with respect to $\sigma$--field $\mathcal{I}_{t-1}$, defined below and depends on random coefficient matrix $\Gamma_{s_t}:=[B_{0,s_t}:B_{1,s_t}:\dots:B_{p_*,s_t}]$. Here $B_{0,s_t}$ is an $(n_*\times k_*)$ random matrix, for $i=1,\dots,p_*$, $B_{i,s_t}$ are $(n_*\times n_*)$ random matrices, and $\varepsilon_1,\dots,\varepsilon_T$ is a random sequence of independent identically multivariate normally distributed random vectors with means of 0 and covariance matrices of $n$ dimensional identity matrix $I_n$. Then, in particular, for multivariate GARCH process of $(0,p_*)$ order, dependence of $\Sigma_t$ on $\Gamma_{s_t}$ is given by 
\begin{equation*}\label{01003}
\text{vech}\big(\Sigma_t\big)=B_{0,s_t}+\sum_{i=1}^{p_*}B_{i,s_t}\text{vech}(\Sigma_{t-i}),
\end{equation*}
where $B_{0,s_t}\in \mathbb{R}^{n(n+1)/2}$ and $B_{i,s_t}\in \mathbb{R}^{[n(n+1)/2]\times [n(n+1)/2]}$ for $i=1,\dots, p_*$ are suitable random vector and matrices and the vech is an operator that stacks elements on and below a main diagonal of a square matrix. 

Let us introduce stacked vectors and matrices: $y:=(y_1',\dots,y_T')'$, $s:=(s_1,\dots,s_T)'$, $\Pi_{\bar{s}_t}:=[\Pi_{s_1}:\dots:\Pi_{s_t}]$, $\Gamma_{\bar{s}_t}:=[\Gamma_{s_1}:\dots:\Gamma_{s_t}]$, $\Pi_s:=[\Pi_{s_1}:\dots:\Pi_{s_T}]$, and $\Gamma_s:=[\Gamma_{s_1}:\dots:\Gamma_{s_T}]$. We also assume that the strong white noise process $\{\varepsilon_t\}_{t=1}^T$ is independent of the random coefficient matrices $\Pi_s$ and $\Gamma_s$, random transition matrix $\mathsf{P}$, and regime--switching vector $s$ conditional on initial information $\mathcal{F}_0:=\sigma(y_{1-p},\dots,y_0,\psi_{1},\dots,\psi_T,\Sigma_{1-p_*},\dots,\Sigma_0)$. Here for a generic random vector $X$, $\sigma(X)$ denotes a $\sigma$--field generated by the random vector $X$, $y_{1-p},\dots,y_0$ are initial values of the process $y_t$, $\Sigma_{1-q_*},\dots,\Sigma_0$ are initial values of the random matrix process $\Sigma_t$, and $\psi_1,\dots,\psi_T$ are exogenous variables and they are known at time zero. We further suppose that the transition probability matrix $\mathsf{P}$ is independent of the random coefficient matrices $\Pi_s$ and $\Gamma_s$ given initial information $\mathcal{F}_0$ and regime--switching vector $s$.

To ease of notations, for a generic vector $o=(o_1,\dots,o_T)'$, we denote its first $t$ and last $T-t$ sub vectors by $\bar{o}_t$ and $\bar{o}_t^c$, respectively, that is, $\bar{o}_t:=(o_1,\dots,o_t)'$ and $\bar{o}_t^c:=(o_{t+1},\dots,o_T)'$. We define $\sigma$--fields: for $t=0,\dots,T$, $\mathcal{F}_{t}:=\mathcal{F}_0\vee\sigma(\bar{y}_{t})$, $\mathcal{H}_t=\mathcal{F}_t\vee \sigma(\Pi_s)\vee \sigma(\Gamma_s)\vee \sigma(\mathsf{P})\vee\sigma(s)$ and for $t=1,\dots,T$, $\mathcal{I}_{t-1}=\mathcal{F}_{t-1}\vee\sigma(\Pi_{\bar{s}_t})\vee\sigma(\Gamma_{\bar{s}_t})\vee \sigma(\mathsf{P})\vee \sigma(\bar{s}_t)$, where for generic sigma fields $\mathcal{O}_1,\dots,\mathcal{O}_k$, $\vee_{i=1}^k \mathcal{O}_i $ is the minimal $\sigma$--field containing the $\sigma$--fields $\mathcal{O}_i$, $i=1,\dots,k$. Observe that $\mathcal{F}_{t}\subset \mathcal{H}_{t}$ for $t=0,\dots,T$. The $\sigma$-fields play major roles in the paper. For the first--order Markov chain, a conditional probability that the regime at time $t+1$, $s_{t+1}$ equals some particular value conditional on the past regimes $\bar{s}_t$, transition probability matrix $\mathsf{P}$, and initial information $\mathcal{F}_0$, depends only through the most recent regime at time $t$, $s_t$, transition probability matrix $\mathsf{P}$, and initial information $\mathcal{F}_0$, that is,
\begin{equation}\label{01004}
p_{s_ts_{t+1}}:=\mathbb{P}[s_{t+1}=s_{t+1}|s_t=s_t,\mathsf{P},\mathcal{F}_0]=\mathbb{P}\big[s_{t+1}=s_{t+1}|\bar{s}_t=\bar{s}_t,\mathsf{P},\mathcal{F}_0\big]
\end{equation} 
for $t=0,\dots,T-1$, where $p_{s_0s_1}=\mathbb{P}[s_1=s_1|\mathsf{P},\mathcal{F}_0]$ is the initial probability. A distribution of a residual random vector $\xi:=(\xi_1',\dots,\xi_T')'$ is given by
\begin{equation*}\label{01006}
\xi=(\xi_1',\dots,\xi_T')'~|~\mathcal{H}_0\sim \mathcal{N}(0,\Sigma),
\end{equation*}
where $\Sigma:=\text{diag}\{\Sigma_1,\dots,\Sigma_T\}$ is a block diagonal matrix. 

To remove duplicates in the random coefficient matrix $(\Pi_s,\Gamma_s)$, for a generic regime--switching vector with length $k$, $o=(o_1,\dots,o_k)'$, we define sets
\begin{equation}\label{08006}
\mathcal{A}_{\bar{o}_t}:=\mathcal{A}_{\bar{o}_{t-1}}\cup\big\{o_t\in \{o_1,\dots,o_k\}\big|o_t\not \in \mathcal{A}_{\bar{o}_{t-1}}\big\},~~~t=1,\dots,k,
\end{equation}
where for $t=1,\dots,k$, $o_t\in \{1,\dots,N\}$ and an initial set is the empty set, i.e., $\mathcal{A}_{\bar{o}_0}=\O$. The final set $\mathcal{A}_o=\mathcal{A}_{\bar{o}_k}$ consists of different regimes in regime vector $o=\bar{o}_k$ and $|\mathcal{A}_o|$ represents a number of different regimes in the regime vector $o$. Let us assume that elements of sets $\mathcal{A}_s$, $\mathcal{A}_{\bar{s}_t}$, and difference sets between the sets $\mathcal{A}_{\bar{s}_t^c}$ and $\mathcal{A}_{\bar{s}_t}$ are given by $\mathcal{A}_s=\{\hat{s}_1,\dots,\hat{s}_{r_{\hat{s}}}\}$, $\mathcal{A}_{\bar{s}_t}=\{\alpha_1,\dots,\alpha_{r_\alpha}\}$, and $\mathcal{A}_{\bar{s}_t^c}\backslash \mathcal{A}_{\bar{s}_t}=\{\delta_1,\dots,\delta_{r_\delta}\}$, respectively, where $r_{\hat{s}}:=|\mathcal{A}_s|$, $r_\alpha:=|\mathcal{A}_{\bar{s}_t}|$, and $r_\delta:=|\mathcal{A}_{\bar{s}_t^c}\backslash \mathcal{A}_{\bar{s}_t}|$ are numbers of elements of the sets, respectively. We introduce the following regime vectors: $\hat{s}:=(\hat{s}_1,\dots,\hat{s}_{r_{\hat{s}}})'$ is an $(r_{\hat{s}}\times 1)$ vector, $\alpha:=(\alpha_1,\dots,\alpha_{r_\alpha})'$ is an $(r_\alpha\times 1)$ vector, and $\delta=(\delta_1,\dots,\delta_{r_\delta})'$ is an $(r_\delta\times 1)$ vector. For the regime vector $a=(a_1,\dots,a_{r_a})' \in\{\hat{s},\alpha,\delta\}$, we also introduce duplication removed random coefficient matrices, whose block matrices are different:  $\Pi_a=[\Pi_{a_1}:\dots:\Pi_{a_{r_a}}]$ is an $(n\times [(np+k)r_a])$ matrix, $\Gamma_a=[\Gamma_{a_1}:\dots:\Gamma_{a_{r_a}}]$ is an $(n_*\times [(n_*(p_*+q_*)+k_*)r_a])$ matrix, and $(\Pi_a,\Gamma_a)$. 

We assume that for given duplication removed regime vector $\hat{s}$ and initial information $\mathcal{F}_0$, the coefficient matrices $(\Pi_{\hat{s}_1},\Gamma_{\hat{s}_1}),\dots,(\Pi_{\hat{s}_{r_{\hat{s}}}},\Gamma_{\hat{s}_{r_{\hat{s}}}})$ are independent under the real probability measure $\mathbb{P}$. Under the assumption, conditional on $\hat{s}$ and $\mathcal{F}_0$, a joint density function of the random coefficient random matrix $(\Pi_{\hat{s}},\Gamma_{\hat{s}})$ is represented by
\begin{equation}\label{08010}
f\big(\Pi_{\hat{s}},\Gamma_{\hat{s}}\big|\hat{s},\mathcal{F}_0\big)=\prod_{t=1}^{r_{\hat{s}}}f\big(\Pi_{\hat{s}_t},\Gamma_{\hat{s}_t}\big|\hat{s}_t,\mathcal{F}_0\big)
\end{equation}
under the real probability measure $\mathbb{P}$, where for a generic random vector $X$, we denote its density function by $f(X)$ under the real probability measure $\mathbb{P}$. Using the regime vectors $\alpha$ and $\delta$, the above joint density function can be written by 
\begin{equation}\label{08011}
f\big(\Pi_{\hat{s}},\Gamma_{\hat{s}}\big|\hat{s},\mathcal{F}_0\big)=
f\big(\Pi_{\alpha},\Gamma_{\alpha}\big|\alpha,\mathcal{F}_0\big)f_*\big(\Pi_{\delta},\Gamma_{\delta}\big|\delta,\mathcal{F}_0\big)
\end{equation}
where the density function $f_*\big(\Pi_{\delta},\Gamma_{\delta}\big|\delta,\mathcal{F}_0\big)$ equals
\begin{equation}\label{08012}
f_*\big(\Pi_\delta,\Gamma_\delta\big|\delta,\mathcal{F}_0\big):=
\begin{cases}
f\big(\Pi_\delta,\Gamma_\delta\big|\delta,\mathcal{F}_0\big),& \text{if}~~~r_\delta\neq 0,\\
1,& \text{if}~~~r_\delta= 0.
\end{cases}
\end{equation}

In order to change from the real probability measure $\mathbb{P}$ to some risk--neutral probability measure $\tilde{\mathbb{P}}$, we define the following state price density process:
\begin{equation*}\label{01007}
L_t:=\prod_{m=1}^t\exp\bigg\{\theta_m'\Sigma_m^{-1}\xi_m-\frac{1}{2}\theta_m'\Sigma_m^{-1}\theta_m\bigg\}
\end{equation*}
for $t=1,\dots,T$, where $\theta_m\in \mathbb{R}^n$ is $\mathcal{I}_{m-1}$ measurable Girsanov kernel process (see, \citeA{Bjork09}) and is defined below. Then it can be shown that $\{L_t\}_{t=0}^T$ is a martingale with respect to a filtration $\{\mathcal{H}_t\}_{t=0}^T$ and the real probability measure $\mathbb{P}$. Therefore, we have $\mathbb{E}[L_T|\mathcal{H}_{0}]=\mathbb{E}[L_1|\mathcal{H}_0]=1$ and $\mathbb{E}[L_T|\mathcal{F}_0]=\mathbb{E}\big[\mathbb{E}[L_T|\mathcal{H}_0]|\mathcal{F}_0\big]=1$. As a result, for all $\omega\in\Omega$, $L_T(\omega)>0$, 
\begin{equation*}\label{01008}
\tilde{\mathbb{P}}\big[A|\mathcal{F}_0\big]=\int_AL_T(\omega|\mathcal{F}_0)d\mathbb{P}\big[\omega|\mathcal{F}_0\big]~~~ \mbox{for all}~A\in \mathcal{H}_T
\end{equation*}
becomes a probability measure, which is called the risk--neutral probability measure. 

By introducing the concept of mean--self--financing, \citeA{Follmer86} extended the concept of the complete market into the incomplete market. In this paper, we will work in the incomplete market. For this reason, we consider a variance of the state price density process 
\begin{equation*}\label{01175}
\text{Var}\big[L_T|\mathcal{F}_0\big]=\bigg\|\frac{d\tilde{\mathbb{P}}}{d\mathbb{P}}-1\bigg\|_2^2
\end{equation*}
where $\|\cdot\|_2$ is a $\mathcal{L}_2$ norm, and relative entropy of the risk--neutral probability measure $\mathbb{\tilde{P}}$ with respect to the real probability measure $\mathbb{P}$, is defined by
\begin{equation*}\label{01176}
I(\mathbb{\tilde{P}},\mathbb{P})=\mathbb{E}\big[L_T\ln(L_T)\big|\mathcal{F}_0\big]=\mathbb{\tilde{E}}\big[\ln(L_T)\big|\mathcal{F}_0\big].
\end{equation*}
Their usage and connection with the incomplete market can be found in \citeA{Frittelli00} and \citeA{Schweizer95}. Let $\bar{\Sigma}_t:=\text{diag}\{\Sigma_1,\dots,\Sigma_t\}$ and $\bar{\Sigma}_t^c:=\text{diag}\{\Sigma_{t+1},\dots,\Sigma_T\}$ be partitions, corresponding to random vectors $\bar{\xi}_t$ and $\bar{\xi}_t^c$ of the covariance matrix $\Sigma$. Then, the following Theorem holds.

\begin{thm}\label{thm01}
Let $\theta=(\theta_1,\dots,\theta_T)'\in\mathbb{R}^{nT}$ be a Girsanov kernel vector, $b=(b_1,\dots,b_q)'\in\mathbb{R}^q$ be random vector, and $\mathcal{A}\in\mathbb{R}^{q\times nT}$ be a full rank random matrix with $q\leq nT$. If we assume that for $t=1,\dots,T$, $\theta_t\in\mathbb{R}^n$, $b$, and $\mathcal{A}$ are $\mathcal{I}_{t-1}$ measurable, then the following results hold
\begin{itemize}
\item[(i)] for $t=0,\dots,T-1$, the following probability laws are true
\begin{equation}\label{01009}
\xi~|~\mathcal{H}_0 \sim \mathcal{N}\big(\theta,\Sigma\big),
\end{equation}
\begin{equation}\label{01010}
\bar{\xi}_t^c~|~\mathcal{H}_t \sim \mathcal{N}\big(\bar{\theta}_t^c,\bar{\Sigma}_t^c\big),
\end{equation}
and
\begin{equation}\label{01011}
\xi_t~|~\mathcal{H}_{t-1} \sim \mathcal{N}\big(\theta_t,\Sigma_t\big)
\end{equation}
under the risk--neutral probability measure $\tilde{\mathbb{P}}$,
\item[(ii)] subject to a constraint $\mathcal{A}\theta=b$,
\begin{equation}\label{01012}
\theta^*=\Sigma\mathcal{A}'\big(\mathcal{A}\Sigma \mathcal{A}'\big)^{-1}b
\end{equation}
is a unique global minimizer of the relative entropy $I(\mathbb{\tilde{P}},\mathbb{P}|\mathcal{F}_0)=\tilde{\mathbb{E}}\big[\ln(L_T)\big|\mathcal{F}_0\big]$ and variance of the state price density $\mathrm{Var}\big[L_T\big|\mathcal{F}_0\big]$. 
\end{itemize}
\end{thm}

Let us divide the Bayesian MS--VAR$(p)$ process $y_t$, namely, 
\begin{equation}\label{01014}
\begin{cases}
z_t:=M_1y_t=M_1\Pi_{s_t}\mathsf{Y}_{t-1}+\zeta_t\\
x_t:=M_2y_t=M_2\Pi_{s_t}\mathsf{Y}_{t-1}+\eta_t
\end{cases},
\end{equation}
where the matrices $M_1:=[I_{n_z}:0]\in\mathbb{R}^{n_z\times n}$ and $M_2:=[0:I_{n_x}]\in\mathbb{R}^{n_x\times n}$ with $n=n_z+n_x$ are used to extract the vectors $z_t$ and $x_t$ from the random process $y_t$ and $\zeta_t:=M_1\xi_t$ and $\eta_t:=M_2\xi_t$ are residual processes, corresponding to the process $z_t$ and $x_t$. In this case, partitions of the covariance matrix $\Sigma_t$ are given by $\Sigma_{11,t}:=M_1\Sigma_tM_1'$, $\Sigma_{12,t}:=M_1\Sigma_tM_2'$, $\Sigma_{21,t}:=M_2\Sigma_tM_1'$, and $\Sigma_{22,t}:=M_2\Sigma_tM_2'$. For a generic square matrix $O$, we denote a vector, consisting of diagonal elements of the matrix $O$ by $\mathcal{D}[O]$. For system \eqref{01014}, the following Corollary holds.

\begin{cor}\label{cor01}
Let for $t=1,\dots,T$, $\hat{\theta}_{2,t}\in\mathbb{R}^{n_x}$ and $R_{2,t}\in\mathbb{R}^{n_x\times n_x}$ be $\mathcal{I}_{t-1}$ measurable given random vector and invertible matrix. Then, the following results hold
\begin{itemize}
\item[(i)] subject to constraints $\tilde{\mathbb{E}}\big[\exp\big\{R_{2,t}(\eta_t-\hat{\theta}_{2,t})\big\}\big|\mathcal{H}_{t-1}\big]=i_{n_x}$ for $t=1,\dots,T$, a Girsanov kernel process 
\begin{equation}\label{01015}
\theta_t^*:=\begin{bmatrix}
\Sigma_{12,t}\Sigma_{22,t}^{-1}(\hat{\theta}_{2,t}-\alpha_{2,t})\\
\hat{\theta}_{2,t}-\alpha_{2,t}
\end{bmatrix}=\Theta_t(\hat{\theta}_{2,t}-\alpha_{2,t}),
~~~t=1,\dots,T
\end{equation}
is a unique global minimizer of variance of the state price density $\mathrm{Var}\big[L_T\big|\mathcal{F}_0\big]$ and the relative entropy $I(\mathbb{\tilde{P}},\mathbb{P}|\mathcal{F}_0)$, where $\alpha_{2,t}:=\frac{1}{2}R_{2,t}^{-1}\mathcal{D}\big[R_{2,t}\Sigma_{22,t}R_{2,t}'\big]\in\mathbb{R}^{n_x\times n_x}$ and $\Theta_t:=\big[(\Sigma_{12,t}\Sigma_{22,t}^{-1})':I_{n_x}\big]'\in\mathbb{R}^{n\times n_x}$,
\item[(ii)] subject to constraints $\tilde{\mathbb{E}}\big[\eta_t-\hat{\theta}_{2,t}\big|\mathcal{H}_{t-1}\big]=0$ for $t=1,\dots,T$, a Girsanov kernel process
\begin{equation}\label{01016}
\theta_t^*:=\begin{bmatrix}
\Sigma_{12,t}\Sigma_{22,t}^{-1}\hat{\theta}_{2,t}\\
\hat{\theta}_{2,t}
\end{bmatrix}=\Theta_t\hat{\theta}_{2,t},
~~~t=1,\dots,T
\end{equation}
is a unique global minimizer of variance of the state price density and the relative entropy,
\item[(iii)] the process $y_t$ is represented by
\begin{equation}\label{01017}
y_t=\Pi_{s_t}\mathsf{Y}_{t-1}+\Theta_t(\hat{\theta}_{2,t}-\alpha_{2,t})+\xi_t,~~~t=1,\dots,T,
\end{equation}
under the risk--neutral risk measure $\mathbb{\tilde{P}}$.
\end{itemize}
\end{cor}

Also, to price options using dividend discount models, one may apply the following Corollary.
\begin{cor}\label{cor02}
Let us assume that the second line of system \eqref{01014} equals $x_t=M_2\Pi_{s_t}\mathsf{Y}_{t-1}+G_t\eta_t$, where $G_t$ is $\mathcal{I}_{t-1}$ measurable invertible matrix. Then, the following results are true
\begin{itemize}
\item[(i)] subject to constraints $\mathbb{\tilde{E}}\big[\exp\{\eta_t-\hat{\theta}_{2,t}\}|\mathcal{H}_{t-1}\big]=i_{n_x}$ for $t=1,\dots,T$, a Girsanov kernel process
\begin{equation}\label{ad03}
\theta_t^*=\Theta_t\bigg(\hat{\theta}_{2,t}-\frac{1}{2}\mathcal{D}[\Sigma_{22,t}]\bigg),~~~t=1,\dots,T,
\end{equation}
is a unique global minimizer of variance of the state price density  and the relative entropy, where $\Theta_t=\big[\big(\Sigma_{12,t}\Sigma_{22,t}^{-1}\big)':G_t'\big]'$.
\item[(ii)] the process $y_t$ is represented by
\begin{equation}\label{ad04}
y_t=\Pi_{s_t}\mathsf{Y}_{t-1}+\Theta_t\bigg(\hat{\theta}_{2,t}-\frac{1}{2}\mathcal{D}[\Sigma_{22,t}]\bigg)+\mathsf{G}_t\xi_t,~~~t=1,\dots,T,
\end{equation}
under the risk--neutral risk measure $\mathbb{\tilde{P}}$, where $\mathsf{G}_t:=\mathrm{diag}\{I_{n_z},G_t\}$ is a block diagonal matrix.
\end{itemize}
\end{cor}

The following notable two Remarks arise from Corollaries 2.1 and 2.2:
\begin{rem}\label{rem01}
If we assume that the residual processes $\zeta_t$ and $\eta_t$ are independent, then since $\Sigma_{12,t}=0$, it follows from equations \eqref{01015}, \eqref{01016}, and \eqref{ad03} that a distribution of the process $z_t$ is same for the real probability measure $\mathbb{P}$ and risk--neutral probability measure $\tilde{\mathbb{P}}$.
\end{rem}
\begin{rem}\label{rem02}
If one models residual processes $\xi_t$ by conditional heteroscedastic models like $\mathrm{ARCH}(q_*)$ and $\mathrm{GARCH}(q_*,p_*)$, then because of the parameters $\Theta$, $\alpha_{2,t}$, and $\mathcal{D}[\Sigma_{22,t}]$, which depend on square terms of $y_{t-1},\dots,y_{t-q_*}$ the optimal Girsanov kernel process can not be linear, which we require in this paper, see equation \eqref{01018}. 
\end{rem}

Consequently, for the rest of the paper, we focus on the heteroscedastic Bayesian MS--VAR process, where the conditional covariance matrix of the residual process does not depend on lagged values of the endogenous process $y_t$. In particular, one can model the conditional covariance matrix by the GARCH($0,p_*$).

We assume that for $t=1,\dots,T$, $\mathcal{I}_{t-1}$ measurable random process $\theta_t\in \mathbb{R}^n$ has the following representation
\begin{equation}\label{01018}
\theta_t=\Delta_{0,t}\psi_t+\Delta_{1,t}y_{t-1}+\dots+\Delta_{p,t}y_{t-p},~~~t=1,\dots,T,
\end{equation}
where $\Delta_{0,t}\in\mathbb{R}^{n\times k}$ and $\Delta_{i,t}\in\mathbb{R}^{n\times n}$, $i = 1,\dots,p$ are $\mathcal{I}_{t-1}$ measurable random coefficient matrices. It should be noted that one can develop option pricing models that correspond to the following Girsanov kernel
$$\tilde{\theta}_t=\Delta_{0,t}\psi_t+\Delta_{1,t}\xi_1+\dots+\Delta_{t-1,t}\xi_{t-1},~~~t=1,\dots,T,$$
where $\Delta_{0,t}\in\mathbb{R}^{n\times k}$ and $\Delta_{i,t}\in\mathbb{R}^{n\times n}$, $i = 1,\dots,t-1$ are $\mathcal{I}_{t-1}$ measurable random coefficient matrices. For Bayesian MS--VARMA process, its Girsanov kernel can be represented by a form like a process $\tilde{\theta}_t$. We refer to option pricing models, corresponding to the process $\tilde{\theta}_t$ as linear option pricing models. Thus, our models, whose Girsanov kernels are given by equation \eqref{01018} are special cases of the linear option pricing models. 

If we define the following matrix and vectors:
$$\Psi:=\begin{bmatrix}
I_n & 0 & \dots & 0 & \dots & 0 & 0\\
-A_{1,s_2}-\Delta_{1,2} & I_n & \dots & 0 & \dots & 0 & 0\\
\vdots & \vdots & \dots & \vdots & \dots & \vdots & \vdots\\
0 & 0 & \dots & -A_{p-1,s_{T-1}}-\Delta_{p-1,T-1} & \dots & I_n & 0\\
0 & 0 & \dots & -A_{p,s_T}-\Delta_{p,T} & \dots & -A_{1,s_T}-\Delta_{1,T} & I_n
\end{bmatrix},$$
and
$$\delta:=\begin{bmatrix}
(A_{0,s_1}+\Delta_{0,1})\psi_1+(A_{1,s_1}+\Delta_{1,1})y_{0}+\dots+(A_{p,s_1}+\Delta_{p,1})y_{1-p}\\ 
(A_{0,s_2}+\Delta_{0,2})\psi_2+(A_{2,s_2}+\Delta_{2,2})y_{0}+\dots+(A_{p,s_2}+\Delta_{p,2})y_{2-p}\\
\vdots\\
(A_{0,s_{T-1}}+\Delta_{0,T-1})\psi_{T-1}\\
(A_{0,s_T}+\Delta_{0,T})\psi_T
\end{bmatrix},$$
then the following Theorem, which is a trigger of options pricing under the Bayesian MS--VAR process and which will be used in the rest of the paper holds.

\begin{thm}\label{thm02}
Let Bayesian MS--VAR($p$) process $y_t$ is given by equations \eqref{01001} or \eqref{01002}, for $t=1,\dots,T$, representation of random vector $\theta_t$, which is $\mathcal{I}_{t-1}$ measurable is given by equation \eqref{01018} and
$$\delta=\begin{bmatrix}
\delta_1\\ \delta_2
\end{bmatrix} ~~~\text{and}~~~
\Psi=\begin{bmatrix}
\Psi_{11} & 0\\
\Psi_{21} & \Psi_{22}
\end{bmatrix}$$
be partitions, corresponding to random sub vectors $\bar{y}_t$ and $\bar{y}_t^c$ of the random vector $y$. Then the following probability laws hold:
\begin{eqnarray}
y~|~\mathcal{H}_0 &\sim& \mathcal{N}\Big(\Psi^{-1}\delta,\Psi^{-1}\Sigma(\Psi^{-1})'\Big), \label{01019}\\
\bar{y}_t~|~\mathcal{H}_0 &\sim& \mathcal{N}\Big(\Psi_{11}^{-1}\delta_1,\Psi_{11}^{-1}\bar{\Sigma}_t(\Psi_{11}^{-1})'\Big), \label{01020}\\
\bar{y}_t^c~|~\mathcal{H}_0 &\sim& \mathcal{N}\Big(C_{21}\delta_1+\Psi_{22}^{-1}\delta_2,C_{21}\bar{\Sigma}_t C_{21}' +\Psi_{22}^{-1}\bar{\Sigma}_t^c(\Psi_{22}^{-1})'\Big), \label{01021}\\
\bar{y}_t^c~|~\mathcal{H}_t &\sim& \mathcal{N}\Big(\Psi_{22}^{-1}\big(\delta_2-\Psi_{21}\bar{y}_t\big),\Psi_{22}^{-1}\bar{\Sigma}_t^c(\Psi_{22}^{-1})'\Big), \label{01022}\\ 
y_t~|~\mathcal{H}_{t-1} &\sim& \mathcal{N}\Big(\Pi_{s_t}\mathsf{Y}_{t-1}+\theta_t,\Sigma_t\Big), \label{01023}
\end{eqnarray}
under the probability measure $\tilde{\mathbb{P}}$, where $C_{21}=-\Psi_{22}^{-1}\Psi_{21}\Psi_{11}^{-1}.$ Also, conditional on the initial information $\mathcal{F}_0$, a distribution of the random vector $\mathrm{vec}(s,\mathsf{P})$ is same for the risk--neutral probability measure $\mathbb{\tilde{P}}$ and the real probability measure $\mathbb{P}$ and for a conditional distribution of the random vector $\mathrm{vec}(\Pi_{\hat{s}},\Gamma_{\hat{s}})$, we have
\begin{equation*}\label{01024}
\tilde{\mathbb{P}}\big[\mathrm{vec}(\Pi_{\hat{s}},\Gamma_{\hat{s}})\in B\big|s,\mathsf{P},\mathcal{F}_0\big]=\mathbb{P}\big[\mathrm{vec}(\Pi_{\hat{s}},\Gamma_{\hat{s}})\in B\big|\hat{s},\mathcal{F}_0\big],
\end{equation*}
where $B\in\mathcal{B}(\mathbb{R}^d)$ with $d:=(np+k)nr_{\hat{s}}+(n_*p_*+k_*)n_*r_{\hat{s}}$ is a Borel set.
\end{thm}

In this paper, we will consider non--dividend paying assets. For dividend--paying option pricing model, based on the dividend discount model, we refer to \citeA{Battulga22b}. Because Bayesian analysis relies on Monte--Carlo simulation, the following Lemma is important. 

\begin{lem}\label{lem01}
Let $(X_1,Y_1),\dots,(X_n,Y_n)$ be independent realizations of random vector $(X,Y)\in \mathbb{R}^{m_1}\times \mathbb{R}^{m_2}$, $h(\cdot,\cdot):\mathbb{R}^{m_1}\times \mathbb{R}^{m_2}\to \mathbb{R}$ be a Borel function, and $h(X,Y)$ be an integrable random variable. We define 
\begin{equation*}\label{01025}
\tau_1:=\frac{1}{n}\sum_{i=1}^nh(X_i,Y_i)~~~\text{and}~~~\tau_2:=\frac{1}{n}\sum_{i=1}^ng(Y_i)
\end{equation*}
where $g(Y):=\mathbb{E}[h(X,Y)|Y]$. Then, the following results hold
$$\mathbb{E}[\tau_1]=\mathbb{E}[\tau_2]=\mathbb{E}[h(X,Y)] ~~~\text{and}~~~\mathrm{Var}(\tau_1)\geq \mathrm{Var}(\tau_2).$$
\end{lem}

The Lemma tells us that the two simulation methods have same expectation but the variance of the 1st simulation method ($\tau_1$) is more than the variance of the 2nd simulation method $(\tau_2)$. As a result, to price options, which will appear in subsequent sections using Monte--Carlo methods, one should use the 2nd method, which is better than the 1st method.

\section{Normal System}

In this section, we price Black--Scholes put and call options on arithmetic weighted price using Theorem 1. We impose weights on all underlying assets at all time periods. Therefore, the options depart from existing options, and choices of the weights give us different types of options. In particular, the options contain European options, Asian options, and basket options (see below). To price the options we assume that economic variables that affect prices of domestic assets are placed on the first $n_z$ components and prices of the domestic assets are placed on the next $n_x$ components of Bayesian MS--VAR$(p)$ process $y_t$, respectively. As before, $M_1=[I_{n_z}:0_{n_z\times n_x}]$ and $M_2=[0_{n_x\times n_z}:I_{n_x}]$ are matrices, which can be used to divide Bayesian MS--VAR process $y_t=(z_t',x_t')'$ into sub vectors of the economic variables and prices of the assets. In this case, a domestic market is given by the following system:
\begin{equation}\label{01026}
\begin{cases}
z_t=\Pi_{1,s_t}\mathsf{Y}_{t-1}+\zeta_t\\
x_t=\Pi_{2,s_t}\mathsf{Y}_{t-1}+\eta_t\\
D_t=\frac{1}{(1+r)^t}
\end{cases},~~~t=1,\dots,T,
\end{equation}
where $r$ is a risk--free interest rate, $D_t$ is a domestic discount process, $z_t=(y_{1,t},\dots,y_{n_z,t})'=M_1y_t$ is a vector of the economic variables, $x_t=(y_{n_z+1,t},\dots,y_{n,t})'=M_2y_t$ is a vector of prices of the domestic assets, $\zeta_t=M_1\xi_t$ is a residual process of the process $z_t$ and $\eta_t=M_2\xi_t$ is a residual process of the process $x_t$, respectively, at time $t$, and $\Pi_{1,s_t}=M_1\Pi_{s_t}$ and $\Pi_{2,s_t}=M_2\Pi_{s_t}$, which correspond to processes $z_t$ and $x_t$, respectively, are partition matrices of the coefficient matrix $\Pi_t$. It is clear that the difference of a discounted price process $D_tx_t$ is given by
\begin{equation*}\label{01027}
D_tx_t-D_{t-1}x_{t-1}=D_t\big(\eta_t-\hat{\theta}_{2,t}\big),
\end{equation*}
where $\hat{\theta}_{2,t}:=M_2\big((1+r)y_{t-1}-\Pi_t\mathsf{Y}_{t-1}\big)$ is $\mathcal{I}_{t-1}$ measurable random process. The process $\hat{\theta}_{2,t}$ has the following representation
\begin{equation*}\label{01028}
\hat{\theta}_{2,t}=\hat{\Delta}_{0,t}\psi_t+\hat{\Delta}_{1,t}y_{t-1}+\dots+\hat{\Delta}_{p,t}y_{t-p},
\end{equation*}
where $\hat{\Delta}_{0,t}:=-M_2A_{0,s_t}$, $\hat{\Delta}_{1,t}:=-M_2\big(A_{1,s_t}-(1+r)I_n\big)$, and for $m=2,\dots,p$, $\hat{\Delta}_{m,t}:=-M_2A_{m,s_t}$. According to the First Fundamental Theorem of asset pricing, we require that the discounted price process $D_tx_t$ is a martingale with respect to the filtration $\{\mathcal{H}_t\}_{t=0}^T$ and some risk--neutral probability measure $\tilde{\mathbb{P}}$. Therefore, the following conditions have to hold
\begin{equation}\label{01029}
\tilde{\mathbb{E}}[\eta_t|\mathcal{H}_{t-1}]=\hat{\theta}_{2,t},~~~t=1,\dots,T,
\end{equation}
where $\mathbb{\tilde{E}}$ denotes an expectation under the risk--neutral probability measure $\tilde{\mathbb{P}}$. 

It is worth mentioning that condition \eqref{01029} corresponds only to the residual process $\eta_t$. Thus, we need to impose a condition on the residual processes $\zeta_t$ under the risk--neutral probability measure. This condition is fulfilled by $\tilde{\mathbb{E}}[f(\zeta_t)|\mathcal{H}_{t-1}]=\hat{\theta}_{1,t}$ for any Borel function $f:\mathbb{R}^{n_z}\to\mathbb{R}^{n_z}$ and $\mathcal{H}_{t-1}$ measurable any random vector $\hat{\theta}_{1,t}\in\mathbb{R}^{n_z}$. Because for any admissible choices of $\hat{\theta}_{1,t}$, condition \eqref{01029} holds, the market is incomplete. But prices of the options, which will be defined below are still consistent with the absence of arbitrage. For this reason, to price the options, we use the optimal Girsanov kernel process $\theta_t$, which minimizes the variance of the state price density process at time $T$ and the relative entropy. According to Corollary 1, the optimal Girsanov kernel process is given by
\begin{equation*}\label{01030}
\theta_t^*:=\Theta_t\hat{\theta}_{2,t}~~~\text{for}~t=1,\dots,T,
\end{equation*}
where $\Theta_t:=\big[(\Sigma_{12,t}\Sigma_{22,t}^{-1})':I_{n_x}\big]'$. Consequently, the representation of the Girsanov kernel process in Theorem 2 is given by
\begin{equation}\label{01031}
\theta_t^*=\Delta_{0,t}\psi_t+\Delta_{1,t}y_{t-1}+\dots+\Delta_{p,t}y_{t-p},~~~t=1,\dots,T,
\end{equation}
where for $m=0,\dots,p$, $\Delta_{m,t}:=\Theta\hat{\Delta}_{m,t}$. It should be noted that if we do not consider economic variables that affect the price process $x_t$ in the normal system, that is, $y_t=x_t$, then the normal system \eqref{01026} becomes complete. Also, in this case, since $\Theta_t=I_{n_x}$, one can model the residual process $\xi_t=\eta_t$ by the conditional heteroscedastic processes, e.g., ARCH and GARCH. Due to Theorem \ref{thm02}, for given $\mathcal{H}_t$, a distribution of the random vector $\bar{y}_t^c$ is given by
\begin{equation*}\label{01032}
\bar{y}_t^c=(y_{t+1}',\dots,y_T')'~|~\mathcal{H}_t\sim \mathcal{N}\big(\mu_{2.1}(\bar{y}_t),\Sigma_{22.1}\big)
\end{equation*}
under the risk--neutral measure $\tilde{\mathbb{P}}$, corresponding to the Girsanov kernel process \eqref{01031}, where $\mu_{2.1}(\bar{y}_t):=\Psi_{22}^{-1}\big(\delta_2-\Psi_{21}\bar{y}_t\big)$ and $\Sigma_{22.1}:=\Psi_{22}^{-1}\bar{\Sigma}_t^c(\Psi_{22}^{-1})'$ are mean vector and covariance matrix of the random vector $\bar{y}_t^c$ given $\mathcal{H}_t$, respectively. Note that since normally distributed random vectors can take negative values, prices of the assets take negative values with positive probability. On the other hand the risk--free rate $r$ is constant. Those two things are the main disadvantages of system \eqref{01026}. 

Let $x:=(x_1',\dots,x_T')'$ be a price vector of the domestic assets. Then, it is clear that $x=(I_T\otimes M_2)y$, where $\otimes$ is the Kronecker product of two matrices. Let $w=(w_1',\dots,w_T')'$ be a weight vector, which corresponds to the price vector $x$ and we define an arithmetic weighted price of the price vector $x$ by
\begin{equation*}\label{01033}
\bar{x}_w:=w'x=w'(I_T\otimes M_2)y.
\end{equation*}
As mentioned above, choices of the weight vectors give us different type options. For example, for the European option on $i$--th asset, the weight vector is $w_t=0$, $t=1,\dots,T-1$ and $w_T=(0,\dots,0,\underset{(i)}{1},0,\dots,0)'$ (that is, for the vector $w_T$, its $i$--th component equals 1 and others are zero), for Asian option on $i$--th asset, the weight vector is $w_t=(0,\dots,0,\underset{(i)}{1/T},0,\dots,0)'$, $t=1,\dots,T$ (that is, for each $t=1,\dots,T$,  $i$--th component of the vector $w_t$ equals $1/T$ and others are zero) and for basket option, the weight vector is $w_t=0$, $t=1,\dots,T-1$. 

In order to obtain a conditional distribution of the arithmetic weighted price $\bar{x}_w$, we rewrite it by
\begin{equation*}\label{01034}
\bar{x}_w=\bar{w}_t'(I_t\otimes M_2)\bar{y}_t+(\bar{w}_t^c)'(I_{T-t}\otimes M_2)\bar{y}_t^c.
\end{equation*}
Therefore, conditional on information $\mathcal{H}_t$ the arithmetic weighted price has the following conditional normal distribution
\begin{equation}\label{01173}
\bar{x}_w ~|~ \mathcal{H}_t \sim \mathcal{N}\big(\mu_{\bar{x}_w}(\bar{y}_t),\sigma_{\bar{x}_w}^2\big)
\end{equation}
under risk--neutral measure $\tilde{\mathbb{P}}$, where $\mu_{\bar{x}_w}(\bar{y}_t):=\bar{w}_t'(I_t\otimes M_2)\bar{y}_t+(\bar{w}_t^c)'(I_{T-t}\otimes M_2)\mu_{2.1}(\bar{y}_t)$ and $\sigma_{\bar{x}_w}^2:=(\bar{w}_t^c)'(I_{T-t}\otimes M_2)\Sigma_{22.1}(I_{T-t}\otimes M_2')\bar{w}_t^c$ are mean and variance of the random variable $\bar{x}_w$ given $\mathcal{H}_t$, respectively. To price Black--Scholes call and put options on the arithmetic weighted price, we need the following Lemma. 

\begin{lem}\label{lem02}
Let $X\sim \mathcal{N}(\mu,\sigma^2)$. Then for all $K\in\mathbb{R}$,
\begin{equation}\label{01035}
\mathbb{E}\big[(X-K)^+\big]=\sigma\Bigg[\phi\bigg(\frac{\mu-K}{\sigma}\bigg)+\bigg(\frac{\mu-K}{\sigma}\bigg)\Phi\bigg(\frac{\mu-K}{\sigma}\bigg)\Bigg]
\end{equation}
and
\begin{equation}\label{01036}
\mathbb{E}\big[(K-X)^+\big]=\sigma\Bigg[\phi\bigg(\frac{K-\mu}{\sigma}\bigg)+\bigg(\frac{K-\mu}{\sigma}\bigg)\Phi\bigg(\frac{K-\mu}{\sigma}\bigg)\Bigg],
\end{equation}
where $\phi(x):=\frac{1}{\sqrt{2\pi}}e^{-x^2/2}$ and $\Phi(x)=\int_{-\infty}^x\frac{1}{\sqrt{2\pi}}e^{-u^2/2}du$ are the density function and cumulative distribution function of the standard normal random variable, and for $x\in\mathbb{R}$, $x^+:=\max\{x,0\}$ is a maximum of $x$ and zero.
\end{lem}

For a generic random vector $X$, let us denote a joint density function of the random vector $X$ by $\tilde{f}(X)$ under the risk--neutral probability measure $\mathbb{P}$ to differentiate the joint density function $f(X)$ under real probability measure $\mathbb{\tilde{P}}$. To price options, which appear in this and the following sections, we use the following Lemma.

\begin{lem}\label{lem04}
Conditional on $\mathcal{F}_t$, a joint density of $\big(\Pi_{\hat{s}},\Gamma_{\hat{s}},s,\mathsf{P}\big)$ is given by
\begin{equation}\label{07042}
\tilde{f}\big(\Pi_{\hat{s}},\Gamma_{\hat{s}},s,\mathsf{P}|\mathcal{F}_t\big)=\frac{\tilde{f}(\bar{y}_t|\Pi_{\alpha},\Gamma_{\alpha},\bar{s}_t,\mathcal{F}_0)f(\Pi_{\hat{s}},\Gamma_{\hat{s}}|\hat{s},\mathcal{F}_0)f(s,\mathsf{P}|\mathcal{F}_0)}{\displaystyle \sum_{\bar{s}_t}\bigg(\int_{\Pi_{\alpha},\Gamma_{\alpha}}\tilde{f}(\bar{y}_t|\Pi_{\alpha},\Gamma_{\alpha},\bar{s}_t,\mathcal{F}_0)f(\Pi_{\alpha},\Gamma_{\alpha}|\alpha,\mathcal{F}_0)d\Pi_\alpha d\Gamma_\alpha\bigg)f(\bar{s}_t|\mathcal{F}_0)}
\end{equation}
for $t=1,\dots,T$, where for $t=1,\dots,T$,
\begin{equation}\label{07043}
\tilde{f}(\bar{y}_t|\Pi_{\alpha},\Gamma_{\alpha},\bar{s}_t,\mathcal{F}_0)=\frac{1}{(2\pi)^{nt/2}|\Sigma_{11}|^{1/2}}\exp\Big\{-\frac{1}{2}\big(\bar{y}_t-\mu_1\big)'\Sigma_{11}^{-1}\big(\bar{y}_t-\mu_1\big)\Big\}
\end{equation}
with $\mu_1:=\Psi_{11}^{-1}\delta_1$ and $\Sigma_{11}:=\Psi_{11}^{-1}\bar{\Sigma}_t(\Psi_{11}^{-1})'$. In particular, we have that
\begin{equation}\label{ad001}
\tilde{f}\big(\Pi_{\hat{s}},\Gamma_{\hat{s}},s|\mathcal{F}_t\big)=\frac{\tilde{f}(\bar{y}_t|\Pi_{\alpha},\Gamma_{\alpha},\bar{s}_t,\mathcal{F}_0)f(\Pi_{\hat{s}},\Gamma_{\hat{s}}|\hat{s},\mathcal{F}_0)f(s|\mathcal{F}_0)}{\displaystyle \sum_{\bar{s}_t}\bigg(\int_{\Pi_{\alpha},\Gamma_{\alpha}}\tilde{f}(\bar{y}_t|\Pi_{\alpha},\Gamma_{\alpha},\bar{s}_t,\mathcal{F}_0)f(\Pi_{\alpha},\Gamma_{\alpha}|\alpha,\mathcal{F}_0)d\Pi_\alpha d\Gamma_\alpha\bigg)f(\bar{s}_t|\mathcal{F}_0)}
\end{equation}
for $t=1,\dots,T$.
\end{lem}

If we denote strike prices of the options by $K$, then from Lemma \ref{lem02} and distribution \eqref{01173}, prices at time $t$ of the Black--Scholes call and put options conditional on $\mathcal{H}_t$ are given by 
\begin{eqnarray*}
C_t(\mathcal{H}_t)&=&\frac{1}{(1+r)^{T-t}}\tilde{\mathbb{E}}\big[\big(\bar{x}_w-K\big)^+\big|\mathcal{H}_t\big]\\
&=&\frac{\sigma_{\bar{x}_w}}{(1+r)^{T-t}}\Bigg[\phi\bigg(\frac{\mu_{\bar{x}_w}(\bar{y}_t)-K}{\sigma_{\bar{x}_w}}\bigg)+\bigg(\frac{\mu_{\bar{x}_w}(\bar{y}_t)-K}{\sigma_{\bar{x}_w}}\bigg)\Phi\bigg(\frac{\mu_{\bar{x}_w}(\bar{y}_t)-K}{\sigma_{\bar{x}_w}}\bigg)\Bigg]
\end{eqnarray*}
and
\begin{eqnarray*}
P_t(\mathcal{H}_t)&=&\frac{1}{(1+r)^{T-t}}\tilde{\mathbb{E}}\big[\big(K-\bar{x}_w\big)^+\big|\mathcal{H}_t\big]\\
&=&\frac{\sigma_{\bar{x}_w}}{(1+r)^{T-t}}\Bigg[\phi\bigg(\frac{K-\mu_{\bar{x}_w}(\bar{y}_t)}{\sigma_{\bar{x}_w}}\bigg)+\bigg(\frac{K-\mu_{\bar{x}_w}(\bar{y}_t)}{\sigma_{\bar{x}_w}}\bigg)\Phi\bigg(\frac{K-\mu_{\bar{x}_w}(\bar{y}_t)}{\sigma_{\bar{x}_w}}\bigg)\Bigg],
\end{eqnarray*}
respectively. Therefore, due to Lemma \ref{lem04} and the tower property of conditional expectation, prices at time $t$ ($t=0,\dots,T-1$) of the Black--Scholes call and put options on the arithmetic weighted price with strike price $K$ and maturity $T$ are obtained as 
$$C_t=\frac{1}{(1+r)^{T-t}}\tilde{\mathbb{E}}\big[\big(\bar{x}_w-K\big)^+\big|\mathcal{F}_t\big]=\sum_s\int_{\Pi_{\hat{s}},\Gamma_{\hat{s}}}C_t(\mathcal{H}_t)\tilde{f}(\Pi_{\hat{s}},\Gamma_{\hat{s}},s|\mathcal{F}_t)d\Pi_{\hat{s}} d\Gamma_{\hat{s}}$$
and
$$P_t=\frac{1}{(1+r)^{T-t}}\tilde{\mathbb{E}}\big[\big(K-\bar{x}_w\big)^+\big|\mathcal{F}_t\big]=\sum_s\int_{\Pi_{\hat{s}},\Gamma_{\hat{s}}}P_t(\mathcal{H}_t)\tilde{f}(\Pi_{\hat{s}},\Gamma_{\hat{s}},s|\mathcal{F}_t)d\Pi_{\hat{s}} d\Gamma_{\hat{s}},$$
respectively. Because in a similar manner we can price other options, which are defined in the following sections using Lemma \ref{lem04}, it is sufficient to price the options for the information $\mathcal{H}_t$. 

It should be noted that if we have a method to generate random realization from the distribution of the random vector $\text{vec}(\bar{y}_t^c,\Pi_{\hat{s}},\Gamma_{\hat{s}},s)$ conditional on $\mathcal{F}_t$, then one can price options by Monte--Carlo simulation methods. To price options by Monte--Carlo methods, for a sufficiently large number $\mathcal{L}$, we need to generate random realizations $V(\ell):=\big(\Pi_{\hat{s}(\ell)}(\ell),\Gamma_{\hat{s}(\ell)}(\ell),s(\ell)\big)$, $\ell=1,\dots,\mathcal{L}$ from $f(\Pi_{\hat{s}},\Gamma_{\hat{s}},s|\mathcal{F}_t)$. Then we substitute them into the $C_t(\mathcal{H}_t)$ obtain call option prices at the realizations $V(\ell)$, namely, $C_t(V(\ell))$, $\ell=1,\dots,\mathcal{L}$. According to the tower property of conditional expectation, we have that $C_t=\tilde{\mathbb{E}}\big[C_t(\mathcal{H}_t)\big|\mathcal{F}_t\big].
$ Then, by the law of large numbers, one can approximate the theoretical theoretical option price $C_t$ by the following average 
\begin{equation*}\label{•}
C_t^2:=\frac{1}{\mathcal{L}}\sum_{\ell=1}^\mathcal{L}C_t(V(\ell))
\end{equation*}
According to Lemma \ref{lem01}, this simulation method is better than the following approximation method, which is based on realizations from $f(\bar{y}_t^c,\Pi_{\hat{s}},\Gamma_{\hat{s}},s|\mathcal{F}_t)$:
\begin{equation*}
C_t^1:=\frac{1}{(1+r)^{T-t}}\bigg[\frac{1}{\mathcal{L}}\sum_{\ell=1}^\mathcal{L}\big(\bar{x}_{w}(\ell)-K\big)^+\bigg].
\end{equation*}
Monte--Carlo methods using the Gibbs sampling algorithm for Bayesian MS--VAR process is proposed by authors. In particular, the Monte--Carlo method of the Bayesian MS--AR($p$) process is provided by \citeA{Albert93}, and its multidimensional versions can be found from \citeA{Krolzig97} and \citeA{Battulga24g}.

\section{Log-normal System}

For the normal system, given by equation \eqref{01026}, as mentioned previously in section 3, there is a positive probability that stock prices take negative values and the spot interest rate takes a constant value, which are undesirable properties for prices of stocks and the spot interest rate, respectively. Therefore, we need a model, where stock prices get positive values and the spot interest rate varies from time to time. For this reason and to extend the normal system, in this section, we will consider a domestic--foreign market, see \citeA{Amin91}, \citeA{Bjork09}, and \citeA{Shreve04}. 

Here we assume that financial variables, which consist of domestic log spot rate, foreign log spot rates, domestic assets, foreign assets, and foreign currencies, and economic variables that affect the financial variables are together placed on Bayesian MS--VAR process $y_t$. To extract the financial variables from the process $y_t$, we introduce the following vectors and matrices:
$\bar{e}_i:=(0,\dots,0,1,0,\dots,0)^T\in \mathbb{R}^n$ is a unit vector, whose $i$--th component is 1 and others are zero, as before the matrix $M_1=\big[I_{n_z}:0_{n_z\times n_x}\big]$ corresponds to $n_z$ economic variables, which includes domestic and foreign log spot rates, a matrix $M_2^d:=\big[0_{n_d\times n_z}:I_{n_d}:0_{n_d\times [n_f+n_q]}\big]$ corresponds to $n_d$ non--dividend paying domestic assets, a matrix $M_2^f:=\big[0_{n_f\times [n_z+n_d]}:I_{n_f}:0_{n_f\times n_q}\big]$ corresponds to $n_f$ non--dividend paying foreign assets, and a matrix $M_2^q:=\big[0_{n_q\times [n_z+n_d+n_f]}:I_{n_q}\big]$ corresponds to $n_q$ foreign currencies, where $n_z$, $n_x$, $n_d$, $n_f$, and $n_q$ will be defined below.

Let $n_q$ be a number of foreign countries, for each $i=1,\dots,n_q$, $r_{i,t}^f$ be a spot interest rate and $\tilde{r}_{i,t}^f:=\ln(1+r_{i,t}^f)$ be a log spot interest rate of $i$--th foreign country, respectively, $r_t^d$ be a domestic spot interest rate, and $\tilde{r}_{t}^d:=\ln(1+r_t^d)$ be log domestic spot interest rate. For the rest of the paper, it should be noted that the tilde of variables means the log of the variables. Since the spot interest rates at time $t$ are known at time $(t-1)$, we can assume that $i$--th foreign log spot rate placed on $(i+1)$--th component and the domestic log spot rate placed on the first component of the process $y_{t-1}$. Which means that $\tilde{r}_{t}^d=\bar{e}_1'y_{t-1}$ and $\tilde{r}_{i,t}^f=\bar{e}_{i+1}'y_{t-1}$, $i=1,\dots,n_q$. Let $n_z\geq n_q+1$ and $z_t:=M_1y_t\in\mathbb{R}^{n_z}$ be a vector that includes the domestic and foreign log spot rates. Since the first $n_q+1$ components of the process $z_t$ correspond to the domestic and foreign log spot rates, we assume that other components of the process $z_t$ correspond to economic variables that affect the financial variables. 

Henceforth, for a generic vector $a\in\mathbb{R}^m$, we will use the following vector notations: $\ln(a):=\big(\ln(a_1),\dots,\ln(a_m)\big)'$ and $\exp(a):=\big(\exp(a_1),\dots,\exp(a_m)\big)'$. Let us suppose that $\tilde{x}_t^d:=\ln(x_t^d)=M_2^dy_t\in\mathbb{R}^{n_d}$ is a log price process of the domestic assets, $\tilde{x}_t^f:=\ln(x_t^f)=M_2^fy_t\in\mathbb{R}^{n_f}$ is a log price process of the foreign assets, $\tilde{x}_t^q:=\ln(x_t^q)=M_2^qy_t\in\mathbb{R}^{n_q}$ is a log currency process of the foreign currencies, $x_t:=\big((x_t^d)',(x_t^f)',(x_t^q)'\big)'\in\mathbb{R}^{n_x}$ is a price process, which consists of prices of the all domestic assets, foreign assets, and foreign currencies, and $\tilde{x}_{t}:=\ln(x_t)\in\mathbb{R}^{n_x}$ is a log price process, where $n_x:=n_d+n_f+n_q$ is a total number of domestic assets, foreign assets, and foreign currencies. 

If we denote the dimension of the domestic--foreign system by $n:=n_z+n_x$, the system is given by the following system:
\begin{equation}\label{01040}
\begin{cases}
z_t=\Pi_{1,s_t}\mathsf{Y}_{t-1}+\zeta_t\\
\tilde{x}_t^d=\Pi_{2,s_t}^d\mathsf{Y}_{t-1}+\eta_t^d,~\tilde{x}_t^f=\Pi_{2,s_t}^f\mathsf{Y}_{t-1}+\eta_t^f,~\tilde{x}_t^q=\Pi_{2,s_t}^q\mathsf{Y}_{t-1}+\eta_t^q\\
D_{t}^d=\exp\{-\tilde{r}_{1}^d-\tilde{r}_{2}^d-\dots-\tilde{r}_{t}^d\}=\prod_{m=1}^t\frac{1}{1+r_m^d},\\
D_{i,t}^f=\exp\{-\tilde{r}_{i,1}^f-\tilde{r}_{i,2}^f-\dots-\tilde{r}_{i,t}^f\}=\prod_{m=1}^t\frac{1}{1+r_{i,m}^f}, ~i=1,\dots,n_q\\
\tilde{r}_{t}^d=\bar{e}_1'y_{t-1}~\text{and}~\tilde{r}_{i,t}^f=\bar{e}_{i+1}'y_{t-1},~i=1,\dots,n_q
\end{cases},~t=1,\dots,T,
\end{equation}
where $D_t^d$ is a domestic discount process, $D_{i,t}^f$ is a discount process of $i$--th foreign country, $\zeta_t:=M_1\xi_t$, $\eta_t^d:=M_2^d\xi_t$, $\eta_t^f:=M_2^f\xi_t$, and $\eta_t^q:=M_2^q\xi_t$ are residual processes, and $\Pi_{1,s_t}:=M_1\Pi_{s_t}$, $\Pi_{2,s_t}^d:=M_2^d\Pi_{s_t}$, $\Pi_{2,s_t}^f:=M_2^f\Pi_{s_t}$, and $\Pi_{2,s_t}^q:=M_2^q\Pi_{s_t}$ are random coefficient matrices of the processes $z_t$, $\tilde{x}_t^d$, $\tilde{x}_t^f$, and $\tilde{x}_t^q$, respectively. For the log price vector of foreign assets  $\tilde{x}_t^f$, we assume that for each $i=1,\dots,n_q$, $n_{i,f}$ represents the number of foreign assets of $i$--th country. Thus, it is clear that the total number of foreign assets $n_f$ equals to sum of the number of foreign assets of all countries, i.e., $n_f=n_{1,f}+\dots+n_{n_q,f}$. 

To keep notations simple, we define the following vectors and matrix: 
\begin{equation*}\label{01041}
X_t:=\big((x_t^d)',(x_t^{f}\odot (Jx_t^{q}))',(\mathcal{M}_t^{f}\odot x_t^{q})'\big)'
\end{equation*}
is a price process, consisting of prices of domestic assets, prices of foreign assets in domestic currency, and prices of foreign money market accounts in domestic currency, $\eta_t:=\big((\eta_t^t)',(\eta_t^f)',(\eta_t^q)'\big)'$ is a residual process of the log price process $\tilde{x}_t$,
\begin{eqnarray*}
\hat{\theta}_{2,t}&:=&\Big(\big(M_2^d(y_{t-1}-\Pi_{s_t}\mathsf{Y}_{t-1})+C_{n_d}^dy_{t-1}\big)',\big(M_2^f(y_{t-1}-\Pi_{s_t}\mathsf{Y}_{t-1})+JC_{n_q}^fy_{t-1}\big)',\\
&&~~\big(M_2^q(y_{t-1}-\Pi_{s_t}\mathsf{Y}_{t-1})+(C_{n_q}^d-C_{n_q}^f)y_{t-1}\big)'\Big)'
\end{eqnarray*} 
is a random process, which is measurable with respect to $\sigma$--field $\mathcal{I}_{t-1}$ and is an ingredient of the Girsanov kernel of the domestic--foreign market, and 
$$R_2:=\begin{bmatrix}
I_{n_d} & 0 & 0\\
0 & I_{n_f} & J\\
0 & 0 & I_{n_q} 
\end{bmatrix}$$
is an $(n_x\times n_x)$ matrix, whose rows play major roles in this section, see below, where $\odot$ is the Hadamard product of two vectors, $\mathcal{M}_t^f:=\big(1/D_{1,t}^f,\dots,1/D_{n_q,t}^f\big)'$ is a process of foreign money market accounts, $C_{m}^d:=i_{m}\bar{e}_{1}'$, $m \in \{n_d,n_q\}$ and $C_{n_q}^f:=[0_{n_q\times 1}: I_{n_q}:0_{n_q\times [n-n_q-1]}]$ matrices are used to extract $m$ times duplicated domestic log spot rate and foreign log spot rates from the process $y_{t-1}$, respectively, and
$$J:=\begin{bmatrix}
i_{n_{1,f}} & 0 & \dots & 0\\
0 & i_{n_{2,f}} & \dots & 0\\
\vdots & \vdots & \ddots & \vdots\\
0 & 0 & \dots & i_{n_{n_q,f}} 
\end{bmatrix}$$
is an $(n_f\times n_q)$ matrix, which is used to convert the prices of foreign assets into domestic currency. Then, for the domestically discounted price process, it can be shown that
\begin{equation}\label{01042}
D_t^dX_t=\big(D_{t-1}^dX_{t-1}\big)\odot\exp\Big(R_2\big(\eta_t-\hat{\theta}_{2,t}\big)\Big)=X_0\odot\exp\bigg\{\sum_{m=1}^tR_2\big(\eta_m-\hat{\theta}_{2,m}\big)\bigg\}.
\end{equation}
To write the random process $\hat{\theta}_{2,t}$ in compact form, let us introduce stacked matrices: $M_2:=\big[(M_2^d)':(M_2^f)':(M_2^q)'\big]'$ and $C:=\big[(C_{n_d}^d)':(JC_{n_q}^f)':(C_{n_q}^d-C_{n_q}^f)'\big]'$. Then, the random process $\hat{\theta}_{2,t}$ can be represented by
\begin{equation*}\label{01043}
\hat{\theta}_{2,t}=M_2\big(y_{t-1}-\Pi_{s_t}\mathsf{Y}_{t-1}\big)+Cy_{t-1}=\hat{\Delta}_{0,t}\psi_t+\hat{\Delta}_{1,t}y_{t-1}+\dots+\hat{\Delta}_{p,t}y_{t-p},
\end{equation*}
where $\hat{\Delta}_{0,t}:=-M_2A_{0,s_t}$, $\hat{\Delta}_{1,t}:=M_2\big(I_n-A_{1,s_t}\big)+C$, and for $m=2,\dots,p$, $\hat{\Delta}_{m,t}:=-M_2A_{m,s_t}$. According to equation \eqref{01042}, as $D_{t-1}^dX_{t-1}$ is $\mathcal{H}_{t-1}$ measurable, in order to the discounted price process $D_t^dX_t$ is a martingale with respect to the filtration $\{\mathcal{H}_t\}_{t=0}^T$ and some risk--neutral probability measure $\tilde{\mathbb{P}}$, we must require that 
\begin{equation}\label{01044}
\tilde{\mathbb{E}}\big[\exp\big\{R_2(\eta_t-\hat{\theta}_{2,t})\big\}|\mathcal{H}_{t-1}\big]=i_{n_x},~~~t=1,\dots,T,
\end{equation}
where $\mathbb{\tilde{E}}$ denotes an expectation under the risk--neutral probability measure $\tilde{\mathbb{P}}$. Like normal system \eqref{01026}, log--normal system \eqref{01040} is also incomplete. It is worth mentioning that if we do not consider economic variables that affect the log price process $\tilde{x}_t$ in the log--normal system, that is, $y_t=\tilde{x}_t$ and spot interest rate is constant, then the log--normal system \eqref{01040} becomes complete. Due to Corollary 1, subject to conditions \eqref{01044}, an optimal Girsanov kernel process that minimizes the variance of the state price density process at time $T$ and the relative entropy is given by
\begin{equation*}\label{01045}
\theta_t^*=\Theta_t(\hat{\theta}_{2,t}-\alpha_{2,t}), ~~~t=1,\dots,T,
\end{equation*}
where $\Theta_t:=\big[(\Sigma_{12,t}\Sigma_{22,t}^{-1})':I_{n_x}\big]'$ and $\alpha_{2,t}:=\frac{1}{2}R_2^{-1}\mathcal{D}\big[R_2\Sigma_{22,t}R_2'\big]$. Let us introduce a matrix $\tilde{R}_2=[0:R_2]\in\mathbb{R}^{n_x\times n}$. Then, since $R_2\hat{\theta}_{2,t}=\tilde{R}_2\Theta_t\hat{\theta}_{2,t}$, $\Theta_t\hat{\theta}_{2,t}=\theta_t^*+\Theta_t\alpha_{2,t}$, and $\tilde{R}_2\Theta_t\alpha_{2,t}=R_2\alpha_{2,t}$, in terms of the processes $\xi_t$, $\theta_t$, and $\alpha_{2,t}$ equation \eqref{01042} can be written as
\begin{equation}\label{01046}
D_t^dX_t=X_0\odot\exp\bigg\{\sum_{m=1}^t\tilde{R}_2\big(\xi_t-\theta_m^*\big)-\sum_{m=1}^tR_2\alpha_{2,m}\bigg\}. 
\end{equation}
We will use this equation to change from the risk--neutral probability measure to other useful probability measures, see subsection 4.2. We denote the first column of a generic matrix $O$ by $(O)_1$ and a matrix, which consists of other columns of the matrix $O$ by $(O)_1^c$. Then, the representation of the Girsanov kernel process in Theorem 2 is given by
\begin{equation}\label{01047}
\theta_t^*=\Delta_{0,t}\psi_t+\Delta_{1,t}y_{t-1}+\dots+\Delta_{p,t}y_{t-p},~~~t=1,\dots,T,
\end{equation}
where $(\Delta_{0,t})_1=\Theta_t\big((\hat{\Delta}_{0,t})_1-\alpha_{2,t}\big)$, $(\Delta_{0,t})_1^c=\Theta_t(\hat{\Delta}_{0,t})_1^c$, and for $m=1,\dots,p$, $\Delta_{m,t}:=\Theta_t\hat{\Delta}_{m,t}$. As a result, due to Theorem \ref{thm02},  conditional on $\mathcal{H}_t$, a distribution of the random vector $\bar{y}_t^c$ is given by
\begin{equation*}\label{01048}
\bar{y}_t^c=(y_{t+1}',\dots,y_T')'~|~\mathcal{H}_t \sim \mathcal{N}\big(\mu_{2.1}(\bar{y}_t),\Sigma_{22.1}\big)
\end{equation*}
under a risk--neutral probability measure $\tilde{\mathbb{P}}$, corresponding to the Girsanov kernel process \eqref{01047}, where $\mu_{2.1}(\bar{y}_t):=\Psi_{22}^{-1}\big(\delta_2-\Psi_{21}\bar{y}_t\big)$ and $\Sigma_{22.1}:=\Psi_{22}^{-1}\bar{\Sigma}_t^c(\Psi_{22}^{-1})'$ are mean vector and covariance matrix of the random vector $\bar{y}_t^c$ given $\mathcal{H}_t$, respectively. 

\subsection{Forward Probability Measure}

According to \citeA{Geman95} (see also books of \citeA{Bjork09}, \citeA{Privault12} and \citeA{Shreve04}), clever change of probability measure leads to a significant reduction in the computational burden of derivative pricing. Therefore, in this subsection, we consider the forward probability measure, which is originated from the risk--neutral probability measure $\mathbb{\tilde{P}}$. In this subsection, using the forward probability measure, we price the European options, Margrabe exchange options, and geometric weighted options. 

The forward measure is frequently used to price options, bonds, and interest rate derivatives. For this reason, we define the following domestic $(t,u)$--forward measure:
\begin{equation*}\label{01049}
\mathbb{\hat{P}}_{t,u}^d\big[A\big|\mathcal{H}_t\big]:=\frac{1}{D_t^dB^d_{t,u}(\mathcal{H}_t)}\int_AD_u^d\mathbb{\tilde{P}}\big[\omega\big|\mathcal{H}_t\big],~~~\text{for all}~A\in \mathcal{H}_T
\end{equation*}
where for given $\mathcal{H}_t$, $B_{t,u}^d(\mathcal{H}_t):=\frac{1}{D_t^d}\mathbb{\tilde{E}}[D_u^d|\mathcal{H}_t]$ is a price at time $t$ of a domestic zero--coupon bond paying 1 (face value) at time $u$. A zero--coupon bond is a bond where the face value is repaid at a fixed maturity date. Prior to the maturity date, the bond makes no payment. 

For the rest of the paper, we assume $0\leq t<u\leq T$. Let us introduce vectors that deal with the risk--free spot interest rates of the domestic and foreign countries: vectors $\gamma_{t,u}^d$ and $\gamma_{t,u}^{i,f}$ ($i=1,\dots,n_q$) are defined by $(\gamma_{t,u}^{d})':=\big[i_{u-t-1}'\otimes \bar{e}_1':0_{1\times [(T-u+1)n]}\big]$ and $(\gamma_{t,u}^{i,f})':=\big[i_{u-t-1}'\otimes \bar{e}_{i+1}':0_{1\times [(T-u+1)n]}\big]$. Then, we have that for $t<u$,
\begin{equation}\label{01050}
\sum_{m=t+1}^u\tilde{r}_{m}^d=\tilde{r}_{t+1}^d+(\gamma_{t,u}^{d})'\bar{y}_t^c
\end{equation}
and
\begin{equation*}\label{01051}
\sum_{m=t+1}^u\tilde{r}_{i,m}^f=\tilde{r}_{i,t+1}^f+(\gamma_{t,u}^{i,f})'\bar{y}_t^c~~~\text{for}~~~i=1,\dots,n_q.
\end{equation*}

According to equation \eqref{01050}, two times of negative exponent of a conditional expectation $\mathbb{\tilde{E}}\Big[\frac{D_u^d}{D_t^d}\Big|\mathcal{H}_t\Big]$ can be represented by
\begin{eqnarray}\label{01052}
&&2\sum_{m=t+1}^u\tilde{r}_{m}^d+\big(\bar{y}_t^c-\mu_{2.1}(\bar{y}_t)\big)'\Sigma_{22.1}^{-1}\big(\bar{y}_t^c-\mu_{2.1}(\bar{y}_t)\big)\nonumber\\
&&=\Big(\bar{y}_t^c-\mu_{2.1}(\bar{y}_t)+\Sigma_{22.1}\gamma_{t,u}^d\Big)'\Sigma_{22.1}^{-1} \Big(\bar{y}_t^c-\mu_{2.1}(\bar{y}_t)+\Sigma_{22.1}\gamma_{t,u}^d\Big)\\
&&+2\Big(\tilde{r}_{t+1}+(\gamma_{t,u}^d)'\mu_{2.1}(\bar{y}_t)\Big)-(\gamma_{t,u}^d)'\Sigma_{22.1}\gamma_{t,u}^d\nonumber.
\end{eqnarray}
Consequently, for given $\mathcal{H}_t$, the price at time $t$ of the domestic zero--coupon bond with maturity $u$ is obtained as
\begin{equation*}\label{01053}
B_{t,u}^d(\mathcal{H}_t)=\exp\big\{a_{t,u}^d(\bar{y}_t)\big\},
\end{equation*}
where
\begin{equation}\label{01055}
a_{t,u}^d(\bar{y}_t):=-\tilde{r}_{t+1}^d-(\gamma_{t,u}^d)'\mu_{2.1}(\bar{y}_t)+\frac{1}{2}(\gamma_{t,u}^d)'\Sigma_{22.1}\gamma_{t,u}^d
\end{equation}
is an exponent of the domestic zero--coupon bond's price given information $\mathcal{H}_t$. The first term of the exponent, which is given by equation \eqref{01052} corresponds to the following conditional normal distribution:
\begin{equation}\label{01056}
\bar{y}_t^c=(y_{t+1}',\dots,y_T')' ~|~\mathcal{H}_t \sim\mathcal{N}\Big(\hat{\mu}_{t,u}^d(\bar{y}_t),\Sigma_{22.1}\Big),
\end{equation}
under the $(t,u)$--forward measure $\mathbb{\hat{P}}_{t,u}^d$, where $\hat{\mu}_{t,u}^d(\bar{y}_t):=\mu_{2.1}(\bar{y}_t)-\Sigma_{22.1}\gamma_{t,u}^d$ is an expectation of the random vector $\bar{y}_t^c$ under the forward measure.
Therefore, we obtain that for all $A\in \mathcal{H}_T$,
\begin{equation}\label{01057}
\mathbb{\tilde{E}}[D_u^d1_A|\mathcal{H}_t]=\begin{cases}
D_t^d\exp\big\{a_{t,u}^d(\bar{y}_t)\big\}\mathcal{N}\big(A,\hat{\mu}_{t,u}^d(\bar{y}_t),\Sigma_{22.1}\big) & \text{if}~~~A\not\in \mathcal{H}_t,\\
D_t^d\exp\big\{a_{t,u}^d(\bar{y}_t)\big\}1_A & \text{if}~~~A\in \mathcal{H}_t,
\end{cases}
\end{equation}
where $\mathcal{N}(\mu,\Sigma,O)$ denotes multivariate normal distribution with mean $\mu$ and covariance matrix $\Sigma$ at a generic event $O\in \mathcal{H}_T$ and for a generic event $B\in\mathcal{H}_T$, $1_B$ is an indicator random variable for the event $B$.

Let $\bar{\tilde{x}}^c:=(\tilde{x}_{t+1}',\dots,\tilde{x}_T')'$ be a log price vector of a price vector $\bar{x}_t^c=(x_{t+1}',\dots,x_T')'$. Then, in terms of the vector $\bar{y}_t^c$, the log price vector is represented by $\bar{\tilde{x}}^c=(I_{T-t}\otimes M_2)\bar{y}_t^c$. Let $\bar{w}_t^c=(w_{t+1}',\dots,w_T')'$ be a weight vector, which corresponds to the price vector $\bar{x}_t^c$ and we define a geometrically weighted price of the price vector $\bar{x}_t^c$
\begin{equation*}\label{01058}
x^w:=\prod_{m=t+1}^T x_m^{w_m}=\exp\bigg\{\sum_{m=t+1}^Tw_m'\tilde{x}_m\bigg\}=\exp\Big\{(\bar{w}_t^c)'(I_{T-t}\otimes M_2)\bar{y}_t^c\Big\},
\end{equation*}
where $x_t^{w_t}=x_{1,t}^{w_{1,t}}x_{2,t}^{w_{2,t}}\dots x_{n_x,t}^{w_{n_x,t}}$. Let us consider exchange options with payoff $\big(w_0x^w-\hat{w}_0x^{\hat{w}}\big)^+$, where $w_0$ and $\hat{w}_0$ are positive real numbers and $\hat{w}$ is a weight vector, corresponding to the random vector $x^{\hat{w}}$. Then, according to the forward measure, conditional on $\mathcal{H}_t$, a price at time $t$ of an exchange option is given by
\begin{eqnarray*}\label{01059}
O_t(\mathcal{H}_t)&:=&\frac{1}{D_t^d}\mathbb{\tilde{E}}\Big[D_T^d\big(w_0x^w-\hat{w}_0x^{\hat{w}}\big)^+\Big|\mathcal{H}_t\Big]=B_{t,T}^d(\mathcal{H}_t)\hat{\mathbb{E}}_{t,T}^d\Big[\big(w_0x^w-\hat{w}_0x^{\hat{w}}\big)^+\Big|\mathcal{H}_t\Big]\nonumber\\
&=&B_{t,T}^d(\mathcal{H}_t)\hat{\mathbb{E}}_{t,T}^d\Big[\Big(e^{\ln(w_0)+(\bar{w}_t^c)'(I_{T-t}\otimes M_2)\bar{y}_t^c}-e^{\ln(\hat{w}_0)+(\bar{\hat{w}}_t^c)'(I_{T-t}\otimes M_2)\bar{y}_t^c}\Big)^+\Big|\mathcal{H}_t\Big],
\end{eqnarray*}
where $\hat{\mathbb{E}}_{t,T}^d$ is an expectation under the ($t,T$)--forward probability measure $\hat{\mathbb{P}}_{t,T}^d$. A Margrabe exchange option gives its owner a right, but not the obligation, to exchange one asset for another asset at a specific point in time. To price Margrabe's exchange option, we need the following Lemma.

\begin{lem}\label{lem05}
Let two--dimensional random vector $X$ has the following normal distribution: 
$$X=\begin{bmatrix}
X_1 \\ X_2
\end{bmatrix}\sim \mathcal{N}\bigg(
\begin{bmatrix}
\mu_1\\ \mu_2
\end{bmatrix},
\begin{bmatrix}
\sigma_1^2 & \sigma_{12}\\
\sigma_{12} & \sigma_2^2
\end{bmatrix}\bigg).$$
Then it holds
\begin{eqnarray*}\label{01060}
&&\Psi(\mu_1,\mu_2,\sigma_1^2,\sigma_2^2,\sigma_{12}):=\mathbb{E}\big[\big(e^{X_1}-e^{X_2}\big)^+\big]\nonumber\\
&& =\exp\Big\{\mu_1+\frac{\sigma_1^2}{2}\Big\} \Phi\bigg(\frac{\mu_1-\mu_2+\sigma_1^2-\sigma_{12}}{\sqrt{\sigma_1^2-2\sigma_{12}+\sigma_2^2}}\bigg)-\exp\Big\{\mu_2+\frac{\sigma_2^2}{2}\Big\} \Phi\bigg(\frac{\mu_1-\mu_2+\sigma_{12}-\sigma_2^2}{\sqrt{\sigma_1^2-2\sigma_{12}+\sigma_2^2}}\bigg),
\end{eqnarray*}
where $\Phi(t):=\int_{-\infty}^t \frac{1}{\sqrt{2\pi}}e^{-u^2/2}du.$
\end{lem}

If we take $Z_1:=\ln(w_0)+(\bar{w}_t^c)'(I_{T-t}\otimes M_2)\bar{y}_t^c$ and $Z_2:=\ln(\hat{w}_0)+(\bar{\hat{w}}_t^c)'(I_{T-t}\otimes M_2)\bar{y}_t^c$ in the above Lemma, then according to the distribution of the random vector $\bar{y}_t^c$, which is given by equation \eqref{01056}, one obtains the parameters of the Lemma:
\begin{equation*}\label{01061}
\mu_1:=\mathbb{E}[Z_1|\mathcal{H}_t]=\ln(w_0)+(\bar{w}_t^c)'(I_{T-t}\otimes M_2)\hat{\mu}_{t,T}^d(\bar{y}_t),
\end{equation*}
\begin{equation*}\label{01062}
\mu_2:=\mathbb{E}[Z_2|\mathcal{H}_t]=\ln(\hat{w}_0)+(\bar{\hat{w}}_t^c)'(I_{T-t}\otimes M_2)\hat{\mu}_{t,T}^d(\bar{y}_t),
\end{equation*}
\begin{equation*}\label{01063}
\sigma_1^2:=\text{Var}[Z_1|\mathcal{H}_t]=(\bar{w}_t^c)'(I_{T-t}\otimes M_2)\Sigma_{22.1}(I_{T-t}\otimes M_2')\bar{w}_t^c,
\end{equation*}
\begin{equation*}\label{01064}
\sigma_2^2:=\text{Var}[Z_2|\mathcal{H}_t]=(\bar{\hat{w}}_t^c)'(I_{T-t}\otimes M_2)\Sigma_{22.1}(I_{T-t}\otimes M_2')\bar{\hat{w}}_t^c,
\end{equation*}
and
\begin{equation*}\label{01065}
\sigma_{12}:=\text{Cov}[Z_1,Z_2|\mathcal{H}_t]=(\bar{w}_t^c)'(I_{T-t}\otimes M_2)\Sigma_{22.1}(I_{T-t}\otimes M_2')\bar{\hat{w}}_t^c.
\end{equation*}
By substituting the parameters into Lemma \ref{lem05}, one obtains price at time $t$ of the Margrabe option given $\mathcal{H}_t$, that is,
\begin{eqnarray*}\label{01066}
O_t(w,\hat{w}|\mathcal{H}_t):=\frac{1}{D_t^d}\mathbb{\tilde{E}}\Big[D_T^d\big(w_0x^w-\hat{w}_0x^{\hat{w}}\big)^+\Big|\mathcal{H}_t\Big]=B_{t,T}^d(\mathcal{H}_t)\Psi\big(\mu_1,\mu_2,\sigma_1^2,\sigma_2^2,\sigma_{12}\big).
\end{eqnarray*}
Let $e_i=(0,\dots,0,1,0,\dots,0)'\in \mathbb{R}^{n_x}$ be an unit vector, whose $i$--th component is one and others are zero. Then, it is clear that $i$--th row of the matrix $R_2$ is obtained by $e_i'R_2$. Now, we list special cases of the Margrabe option, corresponding to the domestic assets, foreign assets, and foreign currencies.

\begin{itemize}

\item[1.] For $i=1,\dots,n_d$ and $u=t+1,\dots,T$, conditional on information $\mathcal{H}_t$, prices at time $t$ of European call and put options on $w_{i,u}^d$ units of $i$--th domestic asset with strike price $K$ and maturity $u$ are given by
\begin{eqnarray*}
C_t(\mathcal{H}_t)=\frac{1}{D_t^d}\mathbb{\tilde{E}}\Big[D_u^d\Big(w_{i,u}^dx_{i,u}^d-K\Big)^+\Big|\mathcal{H}_t\Big]=O_t(w,\hat{w}|\mathcal{H}_t),
\end{eqnarray*}
where weights are $w_0:=w_{i,u}^d$, $\hat{w}_0:=K$, $w_u:=e_i'R_2$, $\hat{w}_u:=0$, and for $m=t+1,\dots,T$ ($m\neq u$), $w_m:=0$ and $\hat{w}_m:=0$, and
\begin{eqnarray*}
P_t(\mathcal{H}_t)=\frac{1}{D_t^d}\mathbb{\tilde{E}}\Big[D_u^d\Big(K-w_{i,u}^dx_{i,u}^d\Big)^+\Big|\mathcal{H}_t\Big]= O_t(w,\hat{w}|\mathcal{H}_t),
\end{eqnarray*}
where weights are $w_0:=K$, $\hat{w}_0:=w_{i,u}^d$, $w_u:=0$, $\hat{w}_u:=e_i'R_2$, and for $m=t+1,\dots,T$ ($m\neq u$), and $w_m:=0$ and $\hat{w}_m:=0$, respectively.
\item[2.] For $i=1,\dots,n_q$, $k=1,\dots,n_{i,f}$, and $u=t+1,\dots,T$, conditional on information $\mathcal{H}_t$ prices at time $t$ of European call and put options on $w_{i_k,u}^f$ units of $i_k$--th foreign asset in domestic currency with strike price $K$ and maturity $u$ are given by the following formulas
\begin{eqnarray*}
C_t(\mathcal{H}_t)=\frac{1}{D_t^d}\mathbb{\tilde{E}}\Big[D_u^d\Big(w_{i_k,u}^fx_{i_k,u}^fx_{i,u}^q-K\Big)^+\Big|\mathcal{H}_t\Big]=O_t(w,\hat{w}|\mathcal{H}_t),
\end{eqnarray*}
where weights are $w_0:=w_{i_k,u}^f$, $\hat{w}_0:=K$, $w_u:=e_{a(i,k)}'R_2$ with $a(i,k):=n_d+\sum_{j=1}^{i-1}n_{j,f}+k$, $\hat{w}_u:=0$, and for $m=t+1,\dots,T$ ($m\neq u$), $w_m:=0$ and $\hat{w}_m:=0$ and subscript $i_k$ represents $k$--th foreign asset of $i$--th country, and
\begin{eqnarray*}
P_t(\mathcal{H}_t)&=&\frac{1}{D_t^d}\mathbb{\tilde{E}}\Big[D_u^d\Big(K-w_{i_k,u}^fx_{i_k,u}^fx_{i,u}^q\Big)^+\Big|\mathcal{H}_t\Big]= O_t(w,\hat{w}|\mathcal{H}_t),
\end{eqnarray*}
where weights are $w_0:=K$, $\hat{w}_0:=w_{i_k,u}^f$, $w_u:=0$, $\hat{w}_u:=e_{a(i,k)}'R_2$, and for $m=t+1,\dots,T$ ($m\neq u$), and $w_m:=0$ and $\hat{w}_m:=0$, respectively.
\item[3.] For $i=1,\dots,n_q$ and $u=t+1,\dots,T$, conditional on information $\mathcal{H}_t$ prices at time $t$ of European call and put options on $w_{i,u}^q$ units of $i$--th foreign currency with strike price $K$ and maturity $u$ are given by the following formulas
\begin{eqnarray*}
C_t(\mathcal{H}_t)&=&\frac{1}{D_t^d}\mathbb{\tilde{E}}\Big[D_u^d\Big(w_{i,u}^qx_{i,u}^q-K\Big)^+\Big|\mathcal{H}_t\Big]=O_t(w,\hat{w}|\mathcal{H}_t),
\end{eqnarray*}
where weights are $w_0:=w_{i,u}^q$, $\hat{w}_0:=K$, $w_u:=e_{a(i)}'R_2$ with $a(i):=n_d+n_f+i$, $\hat{w}_u:=0$, and for $m=t+1,\dots,T$ ($m\neq u$), $w_m:=0$ and $\hat{w}_m:=0$, and
\begin{eqnarray*}
P_t(\mathcal{H}_t)&=&\frac{1}{D_t^d}\mathbb{\tilde{E}}\Big[D_u^d\Big(K-w_{i,u}^qx_{i,u}^q\Big)^+\Big|\mathcal{H}_t\Big]= O_t(w,\hat{w}|\mathcal{H}_t),
\end{eqnarray*}
where weights are $w_0:=K$, $\hat{w}_0:=w_{i,u}^q$, $w_u:=0$, $\hat{w}_u:=e_{a(i)}'R_2$, and for $m=t+1,\dots,T$ ($m\neq u$), and $w_m:=0$ and $\hat{w}_m:=0$, respectively.

\item[4.] For $i,j=1,\dots,n_d$ and $u=t+1,\dots,T$, conditional on information $\mathcal{H}_t$ price at time $t$ of Margrabe option, which has a right to exchange $w_{j,u}^d$ units of $j$--th domestic asset into $w_{i,u}^d$ units of $i$--th domestic asset at time $u$ is given by the following formula
\begin{eqnarray*}
C_t(\mathcal{H}_t)&=&\frac{1}{D_t^d}\mathbb{\tilde{E}}\Big[D_u^d\Big(w_{i,u}^dx_{i,u}^d-w_{j,u}^dx_{j,u}^d\Big)^+\Big|\mathcal{H}_t\Big]=O_t(w,\hat{w}|\mathcal{H}_t),
\end{eqnarray*}
where weights are $w_0:=w_{i,u}^d$, $\hat{w}_0:=w_{j,u}^d$, $w_u:=e_i'R_2$, $\hat{w}_u:=e_j'R_2$, and for $m=t+1,\dots,T$ ($m\neq u$), $w_m:=0$ and $\hat{w}_m:=0$.

\item[5.] For $i=1,\dots,n_d$, $j=1,\dots,n_q$, $k=1,\dots,n_{j,f}$ and $u=t+1,\dots,T$, conditional on information $\mathcal{H}_t$ price at time $t$ of Margrabe option, which has a right to exchange $w_{j_k,u}^f$ units of $k$--th foreign asset of $j$--th foreign country in domestic currency into $w_{i,u}^d$ units of $i$--th domestic asset at time $u$ is given by
\begin{eqnarray*}
C_t(\mathcal{H}_t)&=&\frac{1}{D_t^d}\mathbb{\tilde{E}}\Big[D_u^d\Big(w_{i,u}^dx_{i,u}^d-w_{j_k,u}^fx_{j_k,u}^fx_{j,u}^q\Big)^+\Big|\mathcal{H}_t\Big]=O_t(w,\hat{w}|\mathcal{H}_t),
\end{eqnarray*}
where weights are $w_0:=w_{i,u}^d$, $\hat{w}_0:=w_{j_k,u}^f$, $w_u:=e_i'R_2$, $\hat{w}_u:=e_{a(j,k)}'R_2$, and for $m=t+1,\dots,T$ ($m\neq u$), $w_m:=0$ and $\hat{w}_m:=0$, and conditional on information $\mathcal{H}_t$, price at time $t$ of Margrabe option, which has a right to exchange $w_{i,u}^d$ units of $i$--th domestic asset into $w_{j_k,u}^f$ units of $k$--th foreign asset of $j$--th foreign country in domestic currency at time $u$ is given by
\begin{eqnarray*}
P_t(\mathcal{H}_t)&=&\frac{1}{D_t^d}\mathbb{\tilde{E}}\Big[D_u^d\Big(w_{j_k,u}^fx_{j_k,u}^fx_{j,u}^q-w_{i,u}^dx_{i,u}^d\Big)^+\Big|\mathcal{H}_t\Big]=O_t(w,\hat{w}|\mathcal{H}_t),
\end{eqnarray*}
where weights are $w_0:=w_{j_k,u}^f$, $\hat{w}_0:=w_{i,u}^d$, $w_u:=e_{a(j,k)}'R_2$, $\hat{w}_u:=e_{i}'R_2$, and for $m=t+1,\dots,T$ ($m\neq u$), $w_m:=0$ and $\hat{w}_m:=0$.

\item[6.] For $i=1,\dots,n_d$, $j=1,\dots,n_q$ and $u=t+1,\dots,T$, conditional on information $\mathcal{H}_t$ price at time $t$ of Margrabe option, which has a right to exchange $w_{j,u}^q$ units of $j$--th foreign currency into $w_{i,u}^d$ unit of $i$--th domestic asset at time $u$ is given by
\begin{eqnarray*}
C_t(\mathcal{H}_t)&=&\frac{1}{D_t^d}\mathbb{\tilde{E}}\Big[D_u^d\Big(w_{i,u}^dx_{i,u}^d-w_{j,u}^qx_{j,u}^q\Big)^+\Big|\mathcal{H}_t\Big]=O_t(w,\hat{w}|\mathcal{H}_t),
\end{eqnarray*}
where weights are $w_0:=w_{i,u}^d$, $\hat{w}_0:=w_{j,u}^q$, $w_u:=e_i'R_2$, $\hat{w}_u:=e_{a(j)}'R_2$, and for $m=t+1,\dots,T$ ($m\neq u$), $w_m:=0$ and $\hat{w}_m:=0$, and conditional on information $\mathcal{H}_t$, price at time $t$ of Margrabe option, which has a right to exchange $w_{i,u}^d$ unit of $i$--th domestic asset into $w_{j,u}^q$ units of $j$--th foreign currency at time $s$ is given by
\begin{eqnarray*}
P_t(\mathcal{H}_t)&=&\frac{1}{D_t^d}\mathbb{\tilde{E}}\Big[D_u^d\Big(w_{j,u}^qx_{j,u}^q-w_{i,u}^dx_{i,u}^d\Big)^+\Big|\mathcal{H}_t\Big]=O_t(w,\hat{w}|\mathcal{H}_t),
\end{eqnarray*}
where weights are $w_0:=w_{j,u}^q$, $\hat{w}_0:=w_{i,u}^d$, $w_u:=e_{a(j)}'R_2$, $\hat{w}_u:=e_{i}'R_2$, and for $m=t+1,\dots,T$ ($m\neq u$), $w_m:=0$ and $\hat{w}_m:=0$.

\item[7.] For $i,j=1,\dots,n_q$, $r=1,\dots,n_{i,f}$, $k=1,\dots,n_{j,f}$ and $u=t+1,\dots,T$, conditional on information $\mathcal{H}_t$ price at time $t$ of Margrabe option, which has a right to exchange $w_{j_k,u}^f$ units of $k$--th foreign asset of $j$--th foreign country in domestic currency into $w_{i_r,u}^f$ units of $r$--th foreign asset of $i$--th foreign country in domestic currency at time $u$ is given by
\begin{eqnarray*}
C_t(\mathcal{H}_t)&=&\frac{1}{D_t^d}\mathbb{\tilde{E}}\Big[D_u^d\Big(w_{i_r,u}^fx_{i_r,u}^fx_{i,u}^q-w_{j_k,u}^fx_{j_k,u}^fx_{j,u}^q\Big)^+\Big|\mathcal{H}_t\Big]=O_t(w,\hat{w}|\mathcal{H}_t),
\end{eqnarray*}
where weights are $w_0:=w_{i_r,u}^f$, $\hat{w}_0:=w_{j_k,u}^f$, $w_u:=e_{a(i,r)}'R_2$, $\hat{w}_u:=e_{a(j,k)}'R_2$, and for $m=t+1,\dots,T$ ($m\neq u$), $w_m:=0$ and $\hat{w}_m:=0$.

\item[8.] For $i,j=1,\dots,n_q$, $k=1,\dots,n_{i,f}$ and $u=t+1,\dots,T$, conditional on information $\mathcal{H}_t$ price at time $t$ of Margrabe option, which has a right to exchange $w_{j,u}^q$ units of $j$--th foreign currency into $w_{i_k,u}^f$ units of $k$--th foreign asset of $i$--th foreign country in domestic currency at time $u$ is given by 
\begin{eqnarray*}
C_t(\mathcal{H}_t)&=&\frac{1}{D_t^d}\mathbb{\tilde{E}}\Big[D_u^d\Big(w_{i_k,u}^fx_{i_k,u}^fx_{i,u}^q-w_{j,u}^qx_{j,u}^q\Big)^+\Big|\mathcal{H}_t\Big]=O_t(w,\hat{w}|\mathcal{H}_t),
\end{eqnarray*}
where weights are $w_0:=w_{i_k,u}^f$, $\hat{w}_0:=w_{j,u}^q$, $w_u:=e_{a(i,k)}'R_2$, $\hat{w}_u:=e_{a(j)}'R_2$, and for $m=t+1,\dots,T$ ($m\neq u$), $w_m:=0$ and $\hat{w}_m:=0$, and conditional on information $\mathcal{H}_t$, price at time $t$ of Margrabe option, which has a right to exchange $w_{i_k,u}^f$ units of $k$--th foreign asset of $i$--th foreign country in domestic currency into $w_{j,u}^q$ units of $j$--th foreign currency at time $u$ is given by 
\begin{eqnarray*}
P_t(\mathcal{H}_t)&=&\frac{1}{D_t^d}\mathbb{\tilde{E}}\Big[D_u^d\Big(w_{j,u}^qx_{j,u}^q-w_{i_k,u}^fx_{i_k,u}^fx_{i,u}^q\Big)^+\Big|\mathcal{H}_t\Big]=O_t(w,\hat{w}|\mathcal{H}_t),
\end{eqnarray*}
where weights are $w_0:=w_{j,u}^q$, $\hat{w}_0:=w_{i_k,u}^f$, $w_u:=e_{a(j)}'R_2$, $\hat{w}_u:=e_{a(i,k)}'R_2$, and for $m=t+1,\dots,T$ ($m\neq u$), $w_m:=0$ and $\hat{w}_m:=0$.

\item[9.] For $i,j=1,\dots,n_q$ and $u=t+1,\dots,T$, conditional on information $\mathcal{H}_t$ price at time $t$ of Margrabe option, which has a right to exchange $w_{j,u}^q$ unit of $j$--th foreign currency into $w_{i,u}^q$ unit of $i$--th foreign currency at time $u$ is given by
\begin{eqnarray*}
C_t(\mathcal{H}_t)&=&\frac{1}{D_t^d}\mathbb{\tilde{E}}\Big[D_u^d\Big(w_{i,u}^qx_{i,u}^q-w_{j,u}^qx_{j,u}^q\Big)^+\Big|\mathcal{H}_t\Big]=O_t(w,\hat{w}|\mathcal{H}_t),
\end{eqnarray*}
where weights are $w_0:=w_{i,u}^q$, $\hat{w}_0:=w_{j,u}^q$, $w_u:=e_{a(i)}'R_2$, $\hat{w}_u:=e_{a(j)}'R_2$, and for $m=t+1,\dots,T$ ($m\neq u$), $w_m:=0$ and $\hat{w}_m:=0$.
\end{itemize}

\subsection{Change of Probability Measure}

In this section, we consider some probability measures that are originated from the risk--neutral probability measure $\mathbb{\tilde{P}}$. Using the probability measures, we price a general European option, whose special cases are the European options and Margrabe exchange options. 

Let us define the following map defined on $\sigma$-field $\mathcal{H}_T$:
\begin{equation}\label{01068}
\tilde{\mathbb{P}}_{t,u}\big[A|\mathcal{H}_t\big]:=\Bigg(\int_AD_u^dX_ud\mathbb{\tilde{P}}\big[\omega|\mathcal{H}_t\big]\Bigg)\oslash\big(D_t^dX_t\big),~~~A\in\mathcal{H}_T,
\end{equation}
where $\oslash$ is the element--wise division of two vectors. Because the discounted process $D_t^dX_t$ takes positive values and it is a martingale with respect to the filtration $\{\mathcal{H}_t\}_{t=0}^T$ and the risk--neutral probability measure $\mathbb{\tilde{P}}$, each component of the map becomes a probability measure. Note that if we take $A=\Omega$ in equation \eqref{01068}, then as $D_t^dX_t$ is measurable with respect to $\sigma$--field $\mathcal{H}_t$, we have
\begin{equation}\label{01069}
\mathbb{\tilde{E}}\Big[\big(D_u^dX_u\big)\oslash\big(D_t^dX_t\big)\Big|\mathcal{H}_t\Big]=i_{n_x}.
\end{equation}
We denote each component of the map by
\begin{itemize}
\item[($i$)] for domestic assets,
\begin{equation}\label{01070}
\mathbb{\tilde{P}}_{t,u}^{i,d}\big[A|\mathcal{H}_t\big]:=e_i'\tilde{\mathbb{P}}_{t,u}\big[A|\mathcal{H}_t\big]
\end{equation}
for all $A\in \mathcal{H}_T$ and $i=1,\dots,n_d$,
\item[($ii$)] for foreign assets,
\begin{equation}\label{01071}
\mathbb{\tilde{P}}_{t,u}^{i_k,f}\big[A|\mathcal{H}_t\big]:=e_{a(i,k)}'\tilde{\mathbb{P}}_{t,u}\big[A|\mathcal{H}_t\big]
\end{equation}
for all $A\in \mathcal{H}_T$, $i=1,\dots,n_q$, and $k=1,\dots,n_{i,f}$, where the superscript $i_k$ represents $k$--th foreign asset of $i$--th country,
\item[($iii$)] and foreign currencies,
\begin{equation}\label{01072}
\mathbb{\tilde{P}}_{t,u}^{i,q}\big[A|\mathcal{H}_t\big]:=e_{a(i)}'\tilde{\mathbb{P}}_{t,u}\big[A|\mathcal{H}_t\big]
\end{equation}
for all $A\in \mathcal{H}_T$ and $i=1,\dots,n_q$.
\end{itemize}
According to the equation \eqref{01046}, $i$--th component of equation \eqref{01069} is represented by
\begin{equation}\label{01073}
\mathbb{\tilde{E}}\bigg[e_i'\Big(\big(D_u^dX_u\big)\oslash\big(D_t^dX_t\big)\Big)\bigg|\mathcal{H}_t\bigg]=\mathbb{\tilde{E}}\bigg[\exp\bigg\{\sum_{m=t+1}^ue_i'\tilde{R}_2(\xi_m-\theta_m)-\sum_{m=t+1}^ue_i'R_2\alpha_{2,t}\bigg\}\bigg|\mathcal{H}_t\bigg].
\end{equation}
Before we consider the exponent of the expectation, observe that 
\begin{equation*}\label{01074}
\big(\bar{y}_t^c-\mu_{2.1}(\bar{y}_t)\big)'\Sigma_{22.1}^{-1}\big(\bar{y}_t^c-\mu_{2.1}(\bar{y}_t)\big)=\big(\bar{\xi}_t^c-\bar{\theta}_t^c\big)'(\bar{\Sigma}_t^c)^{-1}\big(\bar{\xi}_t^c-\bar{\theta}_t^c\big).
\end{equation*}
As a result, the exponent of expectation \eqref{01073} is given by
\begin{equation*}\label{01075}
\big(\bar{\xi}_t^c-\bar{\theta}_t^c\big)'(\bar{\Sigma}_t^c)^{-1}\big(\bar{\xi}_t^c-\bar{\theta}_t^c\big)-2\sum_{m=t+1}^ue_i'\tilde{R}_2(\xi_m-\theta_m)+2\sum_{m=t+1}^ue_i'\alpha_{2,t}.
\end{equation*}
Let $i_{t,u}:=(i_{u-t}',0)'\in\mathbb{R}^{T-t}$ be a vector, whose first $(u-t)$ elements are 1 and others are zero. Then, the exponent of the expectation is represented by
\begin{eqnarray}\label{01076}
&&\big(\bar{\xi}_t^c-\bar{\theta}_t^c\big)'(\bar{\Sigma}_t^c)^{-1}\big(\bar{\xi}_t^c-\bar{\theta}_t^c\big)-2(i_{t,u}'\otimes e_i'\tilde{R}_2)(\bar{\xi}_t^c-\bar{\theta}_t^c)+2\sum_{m=t+1}^ue_i'R_2\alpha_{2,t}\nonumber\\
&&=\big(\bar{\xi}_t^c-\bar{\theta}_t^c-\bar{\Sigma}_t^c(i_{t,u}\otimes\tilde{R}_2'e_i)\big)'(\bar{\Sigma}_t^c)^{-1}\big(\bar{\xi}_t^c-\bar{\theta}_t^c-\bar{\Sigma}_t^c(i_{t,u}\otimes\tilde{R}_2'e_i)\big)\\
&&-(i_{t,u}'\otimes e_i'\tilde{R}_2)\bar{\Sigma}_t^c(i_{t,u}\otimes\tilde{R}_2'e_i)+2\sum_{m=t+1}^ue_i'R_2\alpha_{2,t}.\nonumber
\end{eqnarray}
It is clear that the last line of equation \eqref{01076} equals zero, that is, $(i_{t,u}'\otimes e_i'\tilde{R}_2)\bar{\Sigma}_t^c(i_{t,u}\otimes\tilde{R}_2'e_i)=2\sum_{m=t+1}^ue_i'R_2\alpha_{2,t}$. Consequently, one finds that
\begin{eqnarray*}\label{01077}
&&\big(\bar{\xi}_t^c-\bar{\theta}_t^c\big)'(\bar{\Sigma}_t^c)^{-1}\big(\bar{\xi}_t^c-\bar{\theta}_t^c\big)-2(i_{t,u}'\otimes e_i\tilde{R}_2)(\bar{\xi}_t^c-\bar{\theta}_t^c)+2\sum_{m=t+1}^ue_i'R_2\alpha_{2,t}\nonumber\\
&&=\Big(\bar{y}_t^c-\mu_{2.1}(\bar{y}_t)-\Psi_{22}^{-1}\bar{\Sigma}_t^c(i_{t,u}\otimes\tilde{R}_2'e_i)\Big)'\Sigma_{22.1}^{-1}\Big(\bar{y}_t^c-\mu_{2.1}(\bar{y}_t)-\Psi_{22}^{-1}\bar{\Sigma}_t^c(i_{t,u}\otimes\tilde{R}_2'e_i)\Big).
\end{eqnarray*}
Hence, conditional on $\mathcal{H}_t$, a distribution of the random vector $\bar{y}_t^c$ is obtained by the following equation
\begin{equation}\label{01078}
\bar{y}_t^c~|~\mathcal{H}_t\sim \mathcal{N}\Big(\mu_{2.1}(\bar{y}_t)+\Psi_{22}^{-1}\bar{\Sigma}_t^c(i_{t,u}\otimes\tilde{R}_2'e_i),\Sigma_{22.1}\Big)
\end{equation}
under the $i$--th probability measure of the map. 

As a result, from equation \eqref{01078}, we have the following distributions
\begin{itemize}
\item[($i$)] for $i$--th ($i=1,\dots,n_d$) domestic asset,
\begin{equation}\label{01079}
\bar{y}_t^c~|~\mathcal{H}_t\sim \mathcal{N}\Big(\mu_{t,u}^{i,d}(\bar{y}_t),\Sigma_{22.1}\Big)
\end{equation}
under the domestic probability measure $\mathbb{\tilde{P}}_{t,u}^{i,d}$, where $\mu_{t,u}^{i,d}(\bar{y}_t):=\mu_{2.1}(\bar{y}_t)+\Psi_{22}^{-1}\bar{\Sigma}_t^c(i_{t,u}\otimes\tilde{R}_2'e_i)$ is an expectation of the the random vector $\bar{y}_t^c$ given $\mathcal{H}_t$,
\item[($ii$)] for $k$--th ($k=1,\dots,n_{i,f}$) asset of $i$--th ($i=1,\dots,n_d$) foreign country,
\begin{equation}\label{01080}
\bar{y}_t^c~|~\mathcal{H}_t\sim \mathcal{N}\Big(\mu_{t,u}^{i_k,f}(\bar{y}_t),\Sigma_{22.1}\Big)
\end{equation}
under the foreign probability measure $\mathbb{\tilde{P}}_{t,u}^{i_k,f}$, where $\mu_{t,u}^{i_k,f}(\bar{y}_t):=\mu_{2.1}(\bar{y}_t)+\Psi_{22}^{-1}\bar{\Sigma}_t^c(i_{t,u}\otimes\tilde{R}_2'e_{a(i,k)})$ is an expectation of the the random vector $\bar{y}_t^c$ given $\mathcal{H}_t$,
\item[($iii$)] and for $i$--th ($i=1,\dots,n_q$) foreign currency,
\begin{equation}\label{01081}
\bar{y}_t^c~|~\mathcal{H}_t\sim \mathcal{N}\Big(\mu_{t,u}^{i,q}(\bar{y}_t),\Sigma_{22.1}\Big)
\end{equation}
under the currency probability measure $\mathbb{\tilde{P}}_{t,u}^{i,q}$, where $\mu_{t,u}^{i,q}(\bar{y}_t):=\mu_{2.1}(\bar{y}_t)+\Psi_{22}^{-1}\bar{\Sigma}_t^c(i_{t,u}\otimes\tilde{R}_2'e_{a(i)})$ is an expectation of the the random vector $\bar{y}_t^c$ given $\mathcal{H}_t$.
\end{itemize}
It follows from equations \eqref{01079}--\eqref{01081} that for all $\psi\in \{d,f,q\}$, $i=1,\dots,n_\psi$ and $A\in \mathcal{H}_T$,
\begin{equation}\label{01082}
\mathbb{\tilde{P}}_{t,u}^{i,\psi}[A|\mathcal{H}_t]=\mathcal{N}\big(A,\mu_{t,u}^{i,\psi}(\bar{y}_t),\Sigma_{22.1}\big).
\end{equation}

Similarly to the domestic zero--coupon bond formula, for $i=1,\dots,n_q$, one can obtain that conditional on information $\mathcal{H}_t$ price at time $t$ of $i$--th country's zero--coupon bond, which expires at time $u$ is given by
\begin{equation*}\label{01083}
B^{i,f}_{t,u}(\mathcal{H}_t)=\exp\big\{-\tilde{r}_{i,t+1}^f-(\gamma_{t,u}^{i,f})' \mu_{2.1}(\bar{y}_t)+\frac{1}{2}(\gamma_{t,u}^{i,f})'\Sigma_{22.1}\gamma_{t,u}^{i,f}\big\}.
\end{equation*}
In order to price options, which are related to foreign currencies, we need to calculate expectations that have forms $\mathbb{\tilde{E}}_{t,u}^{i,q}[D_{i,u}^{f}1_A|\mathcal{H}_t]$, where $\mathbb{\tilde{E}}_{t,u}^{i,q}$ denotes an expectation under the probability measure $\mathbb{\tilde{P}}_{t,u}^{i,q}$. Similarly to the domestic zero--coupon bond, it can be shown that for all $i=1,\dots,n_q$ and $A\in \mathcal{H}_T$,
\begin{equation}\label{01084}
\mathbb{\tilde{E}}_{t,u}^{i,q}[D_{i,u}^{f}1_A|\mathcal{H}_t]=
\begin{cases}
D_{i,t}^{f}\exp\big\{\tilde{a}_{t,u}^{i,q}(\bar{y}_t)\big\}\mathcal{N}\big(A,\tilde{\mu}_{t,u}^{i,q}(\bar{y}_t),\Sigma_{22.1}\big) & \text{if}~~~A\not \in \mathcal{H}_t\\
D_{i,t}^{f}\exp\big\{\tilde{a}_{t,u}^{i,q}(\bar{y}_t)\big\}1_A & \text{if}~~~A \in \mathcal{H}_t
\end{cases}
\end{equation}
where $\tilde{\mu}_{t,u}^{i,q}(\bar{y}_t):=\mu_{t,u}^{i,q}(\bar{y}_t)-\Sigma_{22.1}\gamma_{t,u}^{i,f}$ is an expectation of the random vector $\bar{y}_t^c$ under the currency measure $\mathbb{\tilde{P}}_{t,u}^{i,q}$, and $\tilde{a}_{t,u}^{i,q}(\bar{y}_t):=-\tilde{r}_{i,t+1}^f-(\gamma_{t,u}^{i,f})' \mu_{t,u}^{i,q}(\bar{y}_t)+\frac{1}{2}(\gamma_{t,u}^{i,f})'\Sigma_{22.1}\gamma_{t,u}^{i,f}.$

To illustrate the usage of probability measure change, for the domestic-foreign market, which is given by system \eqref{01040}, we consider a general call option with the following discounted payoff
\begin{equation*}\label{01085}
H_T:=\Bigg[\sum_{u=t+1}^TD_u^d\Bigg(\sum_{i=1}^{n_d}w_{i,u}^dx_{i,u}^d+\sum_{i=1}^{n_q}\Bigg(\sum_{k=1}^{n_{i,f}}w_{i_k,u}^fx_{i_k,u}^fx_{i,u}^q+w_{i,u}^qx_{i,u}^q\Bigg)\Bigg)-D_v^dK\Bigg]^+
\end{equation*}
for $v\geq t$. If we define the following event 
\begin{equation*}\label{01086}
A:=\Bigg\{\sum_{u=t+1}^TD_u^d\Bigg(\sum_{i=1}^{n_d}w_{i,u}^dx_{i,u}^d+\sum_{i=1}^{n_q}\Bigg(\sum_{k=1}^{n_{i,f}}w_{i_k,u}^fx_{i_k,u}^fx_{i,u}^q+w_{i,u}^qx_{i,u}^q\Bigg)\Bigg)\geq D_v^dK\Bigg\},
\end{equation*}
then it follows from the probability measures, which are given by equations \eqref{01070}--\eqref{01072} that conditional on the information $\mathcal{H}_t$, the price at time $t$ of the general call option is given by
\begin{eqnarray*}\label{01087}
C_t(\mathcal{H}_t)&=&\frac{1}{D_t^d}\mathbb{\tilde{E}}\big[H_T\big|\mathcal{H}_t\big]=\sum_{u=t+1}^T\Bigg[\sum_{i=1}^{n_d}w_{i,u}^dx_{i,t}^d\mathbb{\tilde{P}}_{t,u}^{i,d}\big[A\big|\mathcal{H}_t\big]\nonumber\\
&+&\sum_{i=1}^{n_q}\Bigg(\sum_{k=1}^{n_{i,f}}w_{i_k,u}^fx_{i_k,t}^fx_{i,t}^q\mathbb{\tilde{P}}_{t,u}^{i_k,f}\big[A\big|\mathcal{H}_t\big]+w_{i,u}^qM_{i,t}^fx_{i,t}^q\mathbb{\tilde{E}}_{t,u}^{i,q}\big[D_{i,u}^f1_A\big|\mathcal{H}_t\big]\Bigg)\Bigg]\\
&-&\frac{1}{D_t^d}K\mathbb{\tilde{E}}\big[D_v^d1_A\big|\mathcal{H}_t\big].\nonumber
\end{eqnarray*}
Therefore, according to equations \eqref{01057}, \eqref{01082} and \eqref{01084}, we obtain that for given information $\mathcal{H}_t$, price at time $t$ of the call option is given by
\begin{eqnarray*}\label{01088}
C_t(\mathcal{H}_t)&=&\sum_{u=t+1}^T\Bigg[\sum_{i=1}^{n_d}w_{i,u}^dx_{i,t}^d\mathcal{N}\big(A,\mu_{t,u}^{i,d}(\bar{y}_t),\Sigma_{22.1}\big)\nonumber\\
&+&\sum_{i=1}^{n_q}\sum_{k=1}^{n_{i,f}} w_{i_k,u}^fx_{i_k,t}^fx_{i,t}^q\mathcal{N}\big(A,\mu_{t,u}^{i_k,f}(\bar{y}_t),\Sigma_{22.1}\big)\\
&+&\sum_{i=1}^{n_q}w_{i,u}^qx_{i,t}^q\exp\big\{\tilde{a}_{t,u}^{i,q}(\bar{y}_t)\big\}\mathcal{N}\big(A,\tilde{\mu}_{t,u}^{i,q}(\bar{y}_t),\Sigma_{22.1}\big)\Bigg]\nonumber\\
&-&K\exp\big\{a_{t,v}^d(\bar{y}_t)\big\}\mathcal{N}\big(A,\hat{\mu}_{t,v}^d(\bar{y}_t),\Sigma_{22.1}\big)\nonumber.
\end{eqnarray*}
Similarly, one can obtain a pricing formula for a general European put option. Special cases of the general call and put options are the European options and Margrabe options, which are listed in subsection 4.1.

\section{Term Structure Models}

An interest rate swap is an agreement between two parties, where one party pays a fixed interest rate to another party, to receive back a floating interest rate. For this agreement, it can be shown that a forward swap rate is expressed in terms of zero--coupon bonds. A coupon bond is just a weighted sum of zero--coupon bonds. Therefore, to price forward swap rates, coupon bonds, and other interest rate derivatives including cap and floor, one needs the prices of zero--coupon bonds. Price at time $v$ of a zero--coupon bond paying 1 at time $u$ conditional on $\mathcal{H}_v$ is
$$B_{v,u}(\mathcal{H}_v)=\frac{1}{D_v}\mathbb{\tilde{E}}\big[D_u|\mathcal{H}_v\big],~~~0\leq v\leq u\leq T.$$

To price cap, floor, and coupon bond option, it is a well--known fact that it is sufficient to price caplet, floorlet, and zero--coupon bond option, see \citeA{Bjork09}, \citeA{Privault12} and \citeA{Shreve04}. In this section, therefore, we will consider caplet and floorlet for standard forward rate and forward LIBOR rate, and zero--coupon bond option. The standard method to price the caplet and floorlet is based on the instantaneous forward rate. Forward interest rate contract gives its holder a loan decided at time $v$ over a future period of time $[u_1,u_2]$, where we assume $v\leq u_1<u_2$. The interest rate to be applied to this contract is called a forward rate. The forward rate $f_{v,u_1,u_2}$, contracted at time $v$ for a loan $[u_1,u_2]$ is defined from 
\begin{equation}\label{ad01}
\Big(1+f_{v,u_1,u_2}\Big)^{u_2-u_1}=\frac{B_{v,u_1}}{B_{v,u_2}}.
\end{equation}
Instead of the forward rate $f_{v,u_1,u_2}$, we need a log forward rate $\tilde{f}_{v,u_1,u_2}:=\ln(1+f_{v,u_1,u_2})$. It follows from equation \eqref{ad01}, the log forward rate is obtained by
\begin{equation*}\label{01089}
\tilde{f}_{v,u_1,u_2}=-\frac{1}{u_2-u_1}\Big(\ln(B_{v,u_2})-\ln(B_{v,u_1})\Big).
\end{equation*}
A log instantaneous forward rate is defined by $\tilde{f}_{v,u}:=\tilde{f}_{v,u,u+1}$, corresponding to the log forward rates over one period. In terms of the log instantaneous forward rate, the price at time $v$ of a zero--coupon bond with maturity $u$ is represented by
\begin{equation}\label{01090}
B_{v,u}=\exp\big\{-\tilde{f}_{v,v}-\dots-\tilde{f}_{v,u-1}\big\}.
\end{equation}
The log forward rate can be recovered from the log instantaneous forward rates 
\begin{equation}\label{01091}
\tilde{f}_{v,u_1,u_2}:=\frac{1}{u_2-u_1}\Big(\tilde{f}_{v,u_1}+\dots+\tilde{f}_{v,u_2-1}\Big).
\end{equation}

Since term structure models rely on the log instantaneous forward rates, we assume that they are placed on the first $T$ components of the Bayesian MS--VAR$(p)$ process $y_t$. The rest of the components of the process $y_t$ correspond to economic variables that explain the log instantaneous forward rates. In this section, we concentrate only on the domestic market. As a result, our model is given by the following system:
\begin{equation}\label{01092}
\begin{cases}
y_t=\Pi_{s_t}\mathsf{Y}_{t-1}+\xi_t\\
\tilde{f}_{t,v}=\bar{e}_{v-t+1}'y_t\\
\end{cases},~~~t=1,\dots,T,~v=t,\dots,T+t-1,~\text{and}~T\leq n,
\end{equation}
where $n$ is a dimension of the process $y_t$ and $\bar{e}_i\in\mathbb{R}^n$ is the unit vector. Since we consider the domestic market, we will omit superscript $d$ from notations in this section. Because domestic spot interest rate at time $t+1$ equals $f_{t,t}$, that is, $r_{t+1}=f_{t,t}$, for the system, the first component of the process $y_t$ corresponds to the domestic log spot interest rate $\tilde{r}_{t+1}=\ln(1+r_{t+1})$. 

In the Heat--Jarrow--Morton's (HJM) framework of the term structure of forward interest rates, for fixed time $t$, one needs to eliminate arbitrage opportunities, which come from trading bonds with maturities $u=1,\dots,T$. For this reason, by the First Fundamental Theorem, we have to seek risk--neutral probability measure $\mathbb{\tilde{P}}$, which satisfy the following equations (HJM's no--arbitrage conditions)
\begin{equation}\label{01093}
D_{t-1}B_{t-1,u}=\tilde{\mathbb{E}}[D_tB_{t,u}|\mathcal{H}_{t-1}],~~~t=1,\dots,T-1,~u=t+1,\dots,T
\end{equation}
Since $D_t/D_{t-1}=\exp\{-\tilde{f}_{t-1,t-1}\}$, due to second line of equation \eqref{01092} and equation \eqref{01090}, the above equations are written by
\begin{equation*}\label{01094}
\tilde{\mathbb{E}}\bigg[\exp\bigg\{\sum_{m=t}^{u-1}\bar{e}_{m-t+2}'y_{t-1}-\sum_{m=t}^{u-1}\bar{e}_{m-t+1}'y_t\bigg\}\bigg|\mathcal{H}_{t-1}\bigg]=1
\end{equation*}
for $t=1,\dots,T-1$ and $u=t+1,\dots,T$. The above equations can be written by
\begin{equation}\label{ad02}
\tilde{\mathbb{E}}\bigg[\exp\bigg\{J_{1,T-t}y_{t-1}-J_{2,T-t}\Pi_{s_t}\mathsf{Y}_{t-1}-J_{2,T-t}\xi_t\bigg\}\bigg|\mathcal{H}_{t-1}\bigg]=1,~~~t=1,\dots,T-1,
\end{equation}
where the $([T-t]\times n)$ matrices $J_{1,T-t}$ and $J_{2,T-t}$ are given by
\begin{equation*}\label{•}
J_{1,T-t}:=\begin{bmatrix}
0 & 1 & 0 & \dots & 0 & 0 & \dots & 0\\
0 & 1 & 1 & \dots & 0 & 0 & \dots & 0\\
\vdots & \vdots & \vdots & \ddots & \vdots & \vdots & \ddots & \vdots\\
0 & 1 & 1 & \dots & 1 & 0 & \dots & 0
\end{bmatrix}, ~~~
J_{2,T-t}:=\begin{bmatrix}
1 & 0 & \dots & 0 & 0 & \dots & 0\\
1 & 1 & \dots & 0 & 0 & \dots & 0\\
\vdots & \vdots & \ddots & \vdots & \vdots & \ddots & \vdots\\
1 & 1 & \dots & 1 & 0 & \dots & 0
\end{bmatrix}.
\end{equation*}
Since $\xi_t~|~\mathcal{H}_{t-1}\sim \mathcal{N}(\theta_t,\Sigma_t)$, from equation \eqref{ad02}, we have that for $t=1,\dots,T-1$,
\begin{equation}\label{•}
J_{2,T-t}\theta_t=J_{1,T-t}y_{t-1}-J_{2,T-t}\Pi_{s_t}\mathsf{Y}_{t-1}-\frac{1}{2}\mathcal{D}\big[J_{2,T-t}\Sigma_tJ_{2,T-t}'\big].
\end{equation}
Thus, the matrix $\mathcal{A}$ and vector $b$ in Theorem \ref{thm01} are given by
\begin{equation*}\label{•}
\mathcal{A}=\begin{bmatrix}
J_{2,T-1} & 0 & \dots & 0\\
0 & J_{2,T-2} & \dots & 0\\
\vdots & \vdots & \ddots & \vdots\\
0 & 0 & \dots & J_{2,1}
\end{bmatrix}
\end{equation*}
and
\begin{equation*}\label{•}
b=\begin{bmatrix}
J_{1,T-1}y_0-J_{2,T-1}\Pi_{s_1}\mathsf{Y}_0-\frac{1}{2}\mathcal{D}\big[J_{2,T-1}\Sigma_1J_{2,T-1}'\big]\\
J_{1,T-2}y_1-J_{2,T-2}\Pi_{s_2}\mathsf{Y}_1-\frac{1}{2}\mathcal{D}\big[J_{2,T-2}\Sigma_2J_{2,T-2}'\big]\\
\vdots\\
J_{1,1}y_{T-2}-J_{2,1}\Pi_{s_{T-1}}\mathsf{Y}_{T-2}-\frac{1}{2}\mathcal{D}\big[J_{2,1}\Sigma_{T-1}J_{2,1}'\big]
\end{bmatrix},
\end{equation*}
respectively. Consequently, by Theorem \ref{thm01}, for $t=1,\dots,T-1$, $t$--th sub vector of the optimal Girsanov kernel vector $\theta$, which minimizes the relative entropy and variance of the state price density is
\begin{equation*}\label{•}
\theta_t^*=\Sigma_tJ_{2,T-t}'\Big(J_{2,T-t}\Sigma_tJ_{2,T-t}'\Big)^{-1}\bigg(J_{1,T-t}y_{t-1}-J_{2,T-t}\Pi_{s_t}\mathsf{Y}_{t-1}-\frac{1}{2}\mathcal{D}\big[J_{2,T-t}\Sigma_tJ_{2,T-t}'\big]\bigg).
\end{equation*}
Because $D_t$ is measurable $\mathcal{H}_{t-1}$, $B_{t-1,u}=D_t/D_{t-1}=\exp\{-\tilde{f}_{t-1,t-1}\}$, and $B_{t,t}=1$, for $t=1,\dots,T$ and $u=t$, the constraints \eqref{01093} hold under any risk--neutral probability measure, including the real probability measure $\mathbb{P}$. Consequently, $T$--th sub vector of the optimal Girsanov kernel vector is obtained by $\theta_T^*=0$.

By using the optimal Girsanov kernel vector $\theta^*=((\theta_1^*)',\dots,(\theta_{T-1}^*)',0)'$, one obtains risk--neutral probability measure $\tilde{\mathbb{P}}$ (i.e. Girsanov kernel process), which eliminates arbitrage opportunities come from bonds trading. By Theorem \ref{thm02}, a distribution of the random vector $\bar{y}_t^c$ is given by
\begin{equation*}\label{01174}
\bar{y}_t^c~|~\mathcal{H}_t\sim \mathcal{N}(\mu_{2.1}(\bar{y}_t^c),\Sigma_{22.1}),
\end{equation*}
under the risk--neutral probability measure, corresponding to HJM's no--arbitrage conditions, given by equation \eqref{ad02}, where $\mu_{2.1}(\bar{y}_t):=\Psi_{22}^{-1}(\delta_2-\Psi_{21}\bar{y}_t)$ and $\Sigma_{22.1}:=\Psi_{22}^{-1}\bar{\Sigma}_t^c(\Psi_{22}^{-1})'$ are mean vector and covariance matrix of the random vector $\bar{y}_t^c$ given $\mathcal{H}_t$, respectively. 

Let us denote $(t,u)$--forward probability measure, which is originated from the risk--neutral probability measure that satisfies the HJM's no--arbitrage conditions by $\hat{\mathbb{P}}_{t,u}$ as before. By repeating ideas in subsection 4.1, one obtains distribution of the random vector $\bar{y}_t^c$ for given $\mathcal{H}_t$ under the $(t,u)$--forward probability measure $\hat{\mathbb{P}}_{t,u}$, namely
\begin{equation}\label{ad06}
\bar{y}_t^c~|~\mathcal{H}_t\sim \mathcal{N}(\hat{\mu}_{t,u}(\bar{y}_t^c),\Sigma_{22.1})
\end{equation}
the $(t,u)$--forward probability measure $\hat{\mathbb{P}}_{t,u}$, where $\hat{\mu}_{t,u}(\bar{y}_t^c):=\mu_{2.1}(\bar{y}_t^c)-\Sigma_{22.1}\gamma_{t,u}$. According to the $(t,u_2)$--forward measure, for a forward rate caplet, it holds 
\begin{equation*}\label{01101}
\frac{1}{D_t}\mathbb{\tilde{E}}\Big[D_{u_2}\Big(f_{v,u_1,u_2}-\kappa\Big)^+\Big|\mathcal{H}_t\Big]=\exp\big\{a_{t,u_2}(\bar{y}_t)\big\}\mathbb{\hat{E}}_{t,u_2}\Big[\Big(\exp\{\tilde{f}_{v,u_1,u_2}\}-1-\kappa\Big)^+\Big|\mathcal{H}_t\Big]
\end{equation*}
for $t< v\leq u_1< u_2\leq T$. Let us define a matrix $J_{v|t}:=[0:\dots:0:I_n:0:\dots:0]\in\mathbb{R}^{n\times n[T-t]}$, whose $(v-t)$--th block matrix equals $I_n$ and other blocks are zero. This matrix can be used to extract a vector $y_v$ from the vector $\bar{y}_t^c$. By equations \eqref{01091} and \eqref{ad06} and the second line of system \eqref{01092}, conditional on $\mathcal{H}_t$, a distribution of the log forward rate $\tilde{f}_{v,u_1,u_2}$ is given by
\begin{equation}\label{01102}
\tilde{f}_{v,u_1,u_2}~|~\mathcal{H}_t\sim \mathcal{N}\Big(\hat{\mu}_{\tilde{f}_{v,u_1,u_2}}(\bar{y}_t),\sigma_{\tilde{f}_{v,u_1,u_2}}^2\Big)
\end{equation}
under the $(t,u_2)$--forward measure $\mathbb{\hat{P}}_{t,u_2}$, where
\begin{equation*}\label{01103}
\hat{\mu}_{\tilde{f}_{v,u_1,u_2}}(\bar{y}_t):=\frac{1}{u_2-u_1}\bigg(\sum_{m=u_1}^{u_2-1}\bar{e}_{m-v+1}'\bigg)J_{v|t}\hat{\mu}_{t,u_2}(\bar{y}_t)
\end{equation*}
and
\begin{equation*}\label{01104}
\sigma_{\tilde{f}_{v,u_1,u_2}}^2:=\frac{1}{(u_2-u_1)^2}\bigg(\sum_{m=u_1}^{u_2-1}\bar{e}_{m-v+1}'\bigg)J_{v|t}\Sigma_{22.1}J_{v|t}'\bigg(\sum_{m=u_1}^{u_2-1}\bar{e}_{m-v+1}\bigg)
\end{equation*} 
are mean and variance of the log forward rate $\tilde{f}_{v,u_1,u_2}$ given $\mathcal{H}_t$, respectively. In the following Lemma, we reconsider a main Lemma, which is used to price the Black--Scholes European call and put options when the underlying asset follows geometric Brownian motion.

\begin{lem}\label{lem06}
Let $X\sim \mathcal{N}(\mu,\sigma^2)$. Then for all $K>0$,
\begin{equation}\label{01109}
\mathbb{E}\big[\big(e^X-K\big)^+\big]=\exp\Big\{\mu+\frac{\sigma^2}{2}\Big\}\Phi(d_1)-K\Phi(d_2),
\end{equation}
and
\begin{equation}\label{01110}
\mathbb{E}\big[\big(K-e^X\big)^+\big]=K\Phi(-d_2)-\exp\Big\{\mu+\frac{\sigma^2}{2}\Big\}\Phi(-d_1),
\end{equation}
where $d_1:=\big(\mu+\sigma^2-\ln(K)\big)/\sigma$, $d_2:=d_1-\sigma$ and $\Phi(x)=\int_{-\infty}^x\frac{1}{\sqrt{2\pi}}e^{-u^2/2}du$.
\end{lem}

Thus, it follows from the above Lemma that for the forward rate caplet, it holds
\begin{eqnarray*}\label{01111}
&&\frac{1}{D_t}\mathbb{\tilde{E}}\Big[D_{u_2}\Big(f_{v,u_1,u_2}-\kappa\Big)^+\Big|\mathcal{H}_t\Big]=\exp\big\{a_{t,u_2}(\bar{y}_t)\big\}\\
&&\times\bigg[\exp\Big\{\hat{\mu}_{\tilde{f}_{v,u_1,u_2}}(\bar{y}_t)+\frac{1}{2}\sigma_{\tilde{f}_{v,u_1,u_2}}^2\Big\}\Phi\big(d_1\big)-\big(1+\kappa\big)\Phi\big(d_2\big)\bigg]
\end{eqnarray*}
and for the forward rate floorlet, it holds
\begin{eqnarray*}\label{01111}
&&\frac{1}{D_t}\mathbb{\tilde{E}}\Big[D_{u_2}\Big(\kappa-f_{v,u_1,u_2}\Big)^+\Big|\mathcal{H}_t\Big]=\exp\big\{a_{t,u_2}(\bar{y}_t)\big\}\\
&&\times\bigg[\big(1+\kappa\big)\Phi\big(-d_2\big)-\exp\Big\{\hat{\mu}_{\tilde{f}_{v,u_1,u_2}}(\bar{y}_t)+\frac{1}{2}\sigma_{\tilde{f}_{v,u_1,u_2}}^2\Big\}\Phi\big(-d_1\big)\bigg],
\end{eqnarray*}
where $d_1:=\big(\hat{\mu}_{\tilde{f}_{v,u_1,u_2}}(\bar{y}_t)+\sigma_{\tilde{f}_{t,u_1,u_2}}^2-\ln(1+\kappa)\big)/\sigma_{\tilde{f}_{v,u_1,u_2}}$ and $d_2:=d_1-\sigma_{\tilde{f}_{t,u_1,u_2}}.$

Similarly to the standard forward rate contract, a forward rate contract at time $v$ on the LIBOR market provides its holder an interest rate $L_{v,u,w}$ over the future time period $[u,w]$. However, instead exponential compounding for the forward rate, the forward LIBOR rate applies linear compounding. The forward LIBOR rate contracted at time $v$ for a loan $[u_1,u_2]$ is defined from the following equation
\begin{equation*}\label{ad07}
1+(u_2-u_1)L_{v,u_1,u_2}=\frac{B_{v,u_1}}{B_{v,u_2}},
\end{equation*}
see \citeA{Privault12}. Consequently, the above equation implies that
\begin{equation*}\label{01107}
L_{v,u_1,u_2}:=\frac{1}{u_2-u_1}\bigg(\frac{B_{v,u_1}}{B_{v,u_2}}-1\bigg)=\frac{1}{u_2-u_1}\Big(\exp\big\{(u_2-u_1)\tilde{f}_{v,u_1,u_2}\big\}-1\Big),
\end{equation*}
According to equation \eqref{01102}, we have 
\begin{equation*}\label{01108}
(u_2-u_1)\tilde{f}_{v,u_1,u_2}~|~\mathcal{H}_t\sim \mathcal{N}\Big((u_2-u_1)\hat{\mu}_{\tilde{f}_{v,u_1,u_2}}(\bar{y}_t),(u_2-u_1)^2\sigma_{\tilde{f}_{v,u_1,u_2}}^2\Big)
\end{equation*}
under the ($t,u_2$)--forward probability measure $\mathbb{\hat{P}}_{t,u_2}$. By Lemma \ref{lem06}, we can obtain that conditional on the information $\mathcal{H}_t$, prices at time $t$ of LIBOR rate caplet and floorlet are given by
\begin{eqnarray*}\label{01111}
\frac{1}{D_t}\mathbb{\tilde{E}}\Big[D_{u_2}\Big(L_{v,u_1,u_2}-\kappa\Big)^+\Big|\mathcal{H}_t\Big]&=&\frac{1}{u_2-u_1}\exp\big\{a_{t,u_2}(\bar{y}_t)\big\}\bigg[\exp\Big\{(u_2-u_1)\hat{\mu}_{\tilde{f}_{v,u_1,u_2}}(\bar{y}_t)\\
&+&\frac{1}{2}(u_2-u_1)^2\sigma_{\tilde{f}_{v,u_1,u_2}}^2\Big\}\Phi\big(d_1\big)-\big(1+\kappa(u_2-u_1)\big)\Phi\big(d_2\big)\bigg]
\end{eqnarray*}
and
\begin{eqnarray*}\label{01111}
\frac{1}{D_t}\mathbb{\tilde{E}}\Big[D_{u_2}\Big(L_{v,u_1,u_2}-\kappa\Big)^+\Big|\mathcal{H}_t\Big]&=&\frac{1}{u_2-u_1}\exp\big\{a_{t,u_2}(\bar{y}_t)\big\}\bigg[\big(1+\kappa(u_2-u_1)\big)\Phi\big(-d_2\big)\\
&-&\exp\Big\{(u_2-u_1)\hat{\mu}_{\tilde{f}_{v,u_1,u_2}}(\bar{y}_t)+\frac{1}{2}(u_2-u_1)^2\sigma_{\tilde{f}_{v,u_1,u_2}}^2\Big\}\Phi\big(-d_1\big)\bigg]
\end{eqnarray*}
respectively, where 
$$d_1:=\frac{(u_2-u_1)\hat{\mu}_{\tilde{f}_{v,u_1,u_2}}(\bar{y}_t)+(u_2-u_1)^2\sigma_{\tilde{f}_{t,u_1,u_2}}^2-\ln\big(1+\kappa(u_2-u_1)\big)}{(u_2-u_1)\sigma_{\tilde{f}_{v,u_1,u_2}}},~~d_2:=d_1-(u_2-u_1)\sigma_{\tilde{f}_{v,u_1,u_2}}.$$

Now we consider a zero--coupon bond option. In terms of the standard forward rate, the price at time $v$ of a zero--coupon bond paying 1 at time $u$ can be expressed by 
\begin{equation*}\label{01113}
B_{v,u}=\exp\big\{-(u-v)\tilde{f}_{v,v,u}\big\}.
\end{equation*}
Thanks to equation \eqref{01102}, a distribution of exponent of the zero--coupon bond is given by
\begin{equation*}\label{01114}
-(u-v)\tilde{f}_{v,v,u}~|~\mathcal{H}_t\sim \mathcal{N}\Big(-(u-v)\hat{\mu}_{\tilde{f}_{v,v,u}}(\bar{y}_t),(u-v)^2\sigma_{\tilde{f}_{v,v,u}}^2\Big)
\end{equation*}
under the ($t,v$)--forward probability measure $\mathbb{\hat{P}}_{t,v}$. Thus, analogous to the forward LIBOR rate caplet and floorlet one can obtain that prices at time $t$ of European call and put options on price at time $v$ of the zero--coupon bond with maturity $u$ the following formulas are given by
\begin{eqnarray*}\label{01115}
\frac{1}{D_t}\mathbb{\tilde{E}}\Big[D_v\Big(B_{v,u}-K\Big)^+\Big|\mathcal{H}_t\Big]&=&\exp\big\{a_{t,v}(\bar{y}_t)\big\}\bigg[\exp\Big\{-(u-v)\hat{\mu}_{\tilde{f}_{v,v,u}}(\bar{y}_t)\nonumber\\
&+&\frac{1}{2}(u-v)^2\sigma_{\tilde{f}_{v,v,u}}^2\Big\}\Phi\big(d_1\big)-K\Phi\big(d_2\big)\bigg]
\end{eqnarray*}
and
\begin{eqnarray*}\label{01116}
\frac{1}{D_t}\mathbb{\tilde{E}}\Big[D_v\Big(K-B_{v,u}\Big)^+\Big|\mathcal{H}_t\Big]&=&\exp\big\{a_{t,v}(\bar{y}_t)\big\}\bigg[-\exp\Big\{-(u-v)\hat{\mu}_{\tilde{f}_{v,v,u}}(\bar{y}_t)\nonumber\\
&+&\frac{1}{2}(u-v)^2\sigma_{\tilde{f}_{v,v,u}}^2\Big\}\Phi\big(-d_1\big)+K\Phi\big(-d_2\big)\bigg],
\end{eqnarray*}
respectively, where $K$ is the strike price of the bond options and
$$d_1:=\frac{-(u-v)\hat{\mu}_{\tilde{f}_{v,v,u}}(\bar{y}_t)+(u-v)^2\sigma_{\tilde{f}_{v,v,u}}^2-\ln(K)}{(u-v)\sigma_{\tilde{f}_{v,v,u}}^v},~~~d_2:=d_1-(u-v)\sigma_{\tilde{f}_{v,v,u}}.$$

\section{Conclusion}

Economic variables play important roles in any economic model, and sudden and dramatic changes exist in the financial market and economy. The VAR process is workhorse for economic analysis and forecast, but it entails a risk of over--parametrization. Therefore, in the paper, we introduced the Bayesian MS--VAR process to the option pricing models and obtained pricing formulas using the risk--neutral valuation method. It is assumed that the regime--switching process is generated by a Markov process and the residual process follows a conditional heteroscedastic model. For frequently used options, the paper converted previous option pricing models to our option pricing models under Bayesian MS--VAR process. However, the idea of the paper can be used to convert and develop other options. 

It should be noted that Bayesian MS--VAR process contains a simple VAR process, vector error correction model (VECM), BVAR, and MS--VAR process. To use our proposed model, which is based on Bayesian MS--VAR process, one may apply Monte--Carlo methods. The early Monte--Carlo methods can be found in \citeA{Krolzig97}, while recent new method, which removes duplication in the regime--switching vector can be found in \citeA{Battulga24g}. A Monte--Carlo method for large BVAR process, we refer to \citeA{Banbura10}. For simple MS--VAR process, maximum likelihood methods are provided by \citeA{Hamilton89,Hamilton90,Hamilton94} and \citeA{Krolzig97}. To summarize, the main advantages of the paper are

\begin{itemize}
\item because we consider VAR process, the spot rate is not constant and is explained by its own and
other variables' lagged values,
\item it introduced economic variables, regime--switching, and heteroscedasticity,
\item it introduced the Bayesian method for option valuation, so the model will overcome over--parametrization,
\item valuation of options is easy as compared to previous models with regime-switching,
\item and the model contains simple VAR, VECM, BVAR, and MS--VAR processes.
\end{itemize}

\section{Technical Annex}

\begin{proof}[\textbf{Proof of Theorem 1}]
($i$) Since conditional on $\mathcal{H}_0$, a density function of the random vector $\xi=(\xi_1',\dots,\xi_T')'$ is $f(\xi|\mathcal{F}_0)=c\exp\big\{-\frac{1}{2}\sum_{t=1}^T\xi_t'\Sigma_t^{-1}\xi_t\big\}$, we have 
\begin{eqnarray*}\label{01117}
\tilde{\mathbb{P}}\big[\xi\in B\big|\mathcal{H}_0\big]&=&\int_BL_Tf(\xi|\mathcal{F}_0)d\xi=\int_Bc\exp\bigg\{-\frac{1}{2}\sum_{t=1}^T\big(\xi_t-\theta_t\big)'\Sigma_t^{-1}\big(\xi_t-\theta_t\big)\bigg\}d\xi\nonumber\\
&=&\int_Bc\exp\bigg\{-\frac{1}{2}(\xi-\theta)'\Sigma^{-1}(\xi-\theta)\bigg\}d\xi
\end{eqnarray*}
where $B\in \mathcal{B}(\mathbb{R}^{nT})$ is a Borel set and the normalizing coefficient is $c:=\frac{1}{(2\pi)^{nT/2}\prod_{t=1}^T|\Sigma_t|^{1/2}}$. Thus, equation \eqref{01009} holds. As $\Sigma$ is a block diagonal matrix, from the well--known formula of the conditional distribution of a multivariate random vector, one obtains equations \eqref{01010} and \eqref{01011}.

($ii$) Conditional on $\mathcal{H}_0$, the relative entropy $I(\tilde{\mathbb{P}},\mathbb{P}|\mathcal{H}_0)$ of the risk--neutral probability measure $\mathbb{\tilde{P}}$ with respect to the real probability measure $\mathbb{P}$ is given by
\begin{equation}\label{01118}
I(\mathbb{\tilde{P}},\mathbb{P}|\mathcal{H}_0)=\mathbb{\tilde{E}}\big[\ln(L_T)\big|\mathcal{H}_0\big]=\sum_{t=1}^T\mathbb{\tilde{E}}\big[\theta_t'\Sigma_t^{-1}\xi_t\big|\mathcal{H}_0\big]-\frac{1}{2}\sum_{t=1}^T\mathbb{\tilde{E}}\big[\theta_t'\Sigma_t^{-1}\theta_t\big|\mathcal{H}_0\big].
\end{equation}
For the first term of the right--hand side of equation \eqref{01118}, by the tower property of conditional expectation, we have
\begin{equation*}\label{01119}
\mathbb{\tilde{E}}\big[\theta_t'\Sigma_t^{-1}\xi_t\big|\mathcal{H}_0\big]=\mathbb{\tilde{E}}\big[\theta_t'\Sigma_t^{-1}\theta_t\big|\mathcal{H}_0\big].
\end{equation*}
where use the fact that $\xi_t~|~\mathcal{H}_{t-1}\sim \mathcal{N}(\theta_t,\Sigma_t)$ under the risk--neutral probability measure $\mathbb{\tilde{P}}$. 
As a result, for given initial information $\mathcal{F}_0$, the relative entropy is expressed by
\begin{equation*}\label{01121}
I(\mathbb{\tilde{P}},\mathbb{P}|\mathcal{F}_0)=\tilde{\mathbb{E}}\big[\ln(L_T)\big|\mathcal{F}_0\big]=\frac{1}{2}\sum_{t=1}^T\mathbb{\tilde{E}}\big[\theta_t'\Sigma_t^{-1}\theta_t\big|\mathcal{F}_0\big]=\mathbb{\tilde{E}}\bigg[\frac{1}{2}\theta'\Sigma^{-1}\theta\bigg|\mathcal{F}_0\bigg].
\end{equation*}
Since expectation operator preserves inequality for two random variables, it is sufficient to consider the following quadratic programming problem with equality constraints:
\begin{equation}\label{01122}
\begin{cases}
\frac{1}{2}\theta'\Sigma^{-1}\theta\longrightarrow\min\\
\text{s.t.}~\mathcal{A}\theta=b.
\end{cases}
\end{equation}
Since the inverse of covariance matrix $\Sigma^{-1}$ is positive definite, the quadratic programming problem has a unique global minimizer. The Lagrangian function of the quadratic programming problem is
\begin{equation*}\label{01123}
L(\theta,\mu)=\frac{1}{2}\theta'\Sigma^{-1}\theta-\mu'(\mathcal{A}\theta-b),
\end{equation*}
where $\mu\in\mathbb{R}^q$ is a Lagrangian multiplier. Taking partial derivatives from the Lagrangian function with respect to $\theta$ and $\mu$ and setting these partial derivatives to zero, one obtains equation \eqref{01012}.

On the other hand, according to the tower property of conditional expectation, conditional on $\mathcal{H}_0$, the variance of the state price density $L_T$ equals
\begin{equation}\label{•}
\text{Var}\big[L_T\big|\mathcal{H}_0\big]=\mathbb{E}\big[\exp\{\theta'\Sigma^{-1}\theta\}\big|\mathcal{H}_0\big]-1.
\end{equation}
Again, by the tower property of the conditional expectation, we have that
\begin{equation}\label{•}
\text{Var}\big[L_T\big|\mathcal{F}_0\big]=\mathbb{E}\big[\exp\{\theta'\Sigma^{-1}\theta\}\big|\mathcal{F}_0\big]-1.
\end{equation}
Since exponential function is a strictly increasing function, we arrive the optimization problem \eqref{01122}. Thus, a unique global minimizers of the relative entropy and variance of the state price density are identical. That completes the proof of the Theorem.
\end{proof}

\begin{proof}[\textbf{Proof of Corollary 2.1}]
($i$) According to Theorem 1, we have
\begin{equation*}
\xi_t~|~\mathcal{H}_{t-1}\sim \mathcal{N}(\theta_t,\Sigma_t)
\end{equation*}
under the risk--neutral probability measure $\mathbb{\tilde{P}}$. Consequently, for each $t=1,\dots,T$, the constraint 
\begin{equation}\label{01137}
\tilde{\mathbb{E}}\big[\exp\big\{R_{2,t}(\eta_t-\hat{\theta}_{2,t})\big\}\big|\mathcal{H}_{t-1}\big]=\tilde{\mathbb{E}}\big[\exp\big\{\tilde{R}_{2,t}\xi_t-R_{2,t}\hat{\theta}_{2,t})\big\}\big|\mathcal{H}_{t-1}\big]=i_{n_x}
\end{equation}
is equivalent to
\begin{equation*}\label{01138}
\tilde{R}_{2,t}\theta_t=R_{2,t}\hat{\theta}_{2,t}-\frac{1}{2}\mathcal{D}\big[\tilde{R}_{2,t}\Sigma_t\tilde{R}_{2,t}'\big],
\end{equation*} 
where $\tilde{R}_{2,t}:=[0:R_{2,t}]\in\mathbb{R}^{n_z\times n}$. As a result, the matrix $\mathcal{A}$ and the vector $b$ in Theorem \ref{thm01} is given by
\begin{equation*}\label{01139}
\mathcal{A}=\begin{bmatrix}
\tilde{R}_{2,1} & 0 & \dots & 0\\
0 & \tilde{R}_{2,2} & \dots & 0\\
\vdots & \vdots & \ddots & \vdots\\
0 & 0 & \dots & \tilde{R}_{2,T}
\end{bmatrix}~~~\text{and}~~~b=\begin{bmatrix}
R_{2,1}\hat{\theta}_{2,1}-\frac{1}{2}\mathcal{D}\big[\tilde{R}_{2,1}\Sigma_t\tilde{R}_{2,1}'\big]\\
R_{2,2}\hat{\theta}_{2,2}-\frac{1}{2}\mathcal{D}\big[\tilde{R}_{2,2}\Sigma_t\tilde{R}_{2,2}'\big]\\
\vdots\\
R_{2,T}\hat{\theta}_{2,T}-\frac{1}{2}\mathcal{D}\big[\tilde{R}_{2,T}\Sigma_t\tilde{R}_{2,T}'\big]
\end{bmatrix}.
\end{equation*}
According to the optimal Girsanov kernel vector equation \eqref{01012}, corresponding to the relative entropy or variance of state price density process at time $T$, for $t=1,\dots,T$, its $t$--th block vector is represented by
\begin{equation*}\label{01140}
\theta_t^*=\Sigma_t\tilde{R}_{2,t}'\Big(\tilde{R}_{2,t}\Sigma_t\tilde{R}_{2,t}'\Big)^{-1}\bigg(R_{2,t}\hat{\theta}_{2,t}-\frac{1}{2}\mathcal{D}\big[\tilde{R}_{2,t}\Sigma_t\tilde{R}_{2,t}'\big]\bigg).
\end{equation*}
As $\tilde{R}_{2,t}\Sigma_t\tilde{R}_{2,t}'=R_{2,t}\Sigma_{22,t}R_{2,t}'$ and $\big(\tilde{R}_{2,t}\Sigma_t\tilde{R}_{2,t}'\big)^{-1}=(R_{2,t}')^{-1}\Sigma_{22,t}^{-1}R_{2,t}^{-1}$, we obtain equation \eqref{01015}. 

$(ii)$ For each $t=1,\dots,T$, the constraint 
\begin{equation*}\label{01142}
\tilde{\mathbb{E}}\big[\eta_t-\hat{\theta}_{2,t}\big|\mathcal{H}_{t-1}\big]=\tilde{\mathbb{E}}\big[M_2\xi_t-\hat{\theta}_{2,t}\big|\mathcal{H}_{t-1}\big]=0
\end{equation*}
is equivalent to 
\begin{equation*}\label{01143}
M_2\theta_t=\hat{\theta}_{2,t}.
\end{equation*}
As a result, the matrix $\mathcal{A}$ and the vector $b$ in Theorem \ref{thm01} is given by
\begin{equation*}\label{01144}
\mathcal{A}=\begin{bmatrix}
M_2 & 0 & \dots & 0\\
0 & M_2 & \dots & 0\\
\vdots & \vdots & \ddots & \vdots\\
0 & 0 & \dots & M_2
\end{bmatrix}~~~\text{and}~~~b=\begin{bmatrix}
\hat{\theta}_{2,1}\\
\hat{\theta}_{2,2}\\
\vdots\\
\hat{\theta}_{2,T}\\
\end{bmatrix}.
\end{equation*}
Due to the optimal Girsanov kernel vector equation \eqref{01012}, for $t=1,\dots,T$, its $t$--th block vector is represented by
\begin{equation*}\label{01145}
\theta_t^{\text{re}}=\Sigma_tM_2'\big(M_2\Sigma_tM_2'\big)^{-1}\hat{\theta}_{2,t}=\begin{bmatrix}
\Sigma_{12,t}\Sigma_{22,t}^{-1}\hat{\theta}_{2,t}\\
\hat{\theta}_{2,t}
\end{bmatrix}
\end{equation*}
Thus, equation \eqref{01016} holds.

($iii$) Because conditional on $\mathcal{H}_0$, a density function of the random vector $y=(y_1',\dots,y_T')'$ is $f(y|\mathcal{F}_0)=f(\xi|\mathcal{F}_0)=c\exp\big\{-\frac{1}{2}\sum_{t=1}^T(y_t-\Pi_t\mathsf{Y}_{t-1})'\Sigma_t^{-1}(y_t-\Pi_t\mathsf{Y}_{t-1})\big\}$, one obtains that
\begin{eqnarray}\label{01146}
&&\tilde{\mathbb{P}}\big[y\in B\big|\mathcal{H}_0\big]=\int_BL_Tf(y|\mathcal{H}_0)dy\nonumber\\
&&=\int_Bc\exp\bigg\{-\frac{1}{2}\sum_{t=1}^T\big(y_t-(\Pi_{s_t}\mathsf{Y}_{t-1}+\theta_t)\big)'\Sigma_t^{-1}\big(y_t-(\Pi_{s_t}\mathsf{Y}_{t-1}+\theta_t)\big)\bigg\}dy
\end{eqnarray}
where $B\in \mathcal{B}(\mathbb{R}^{nT})$ is a Borel set and the normalizing coefficient is $c:=\frac{1}{(2\pi)^{nT/2}\prod_{t=1}^T|\Sigma_t|^{1/2}}$. Thus, equation \eqref{01017} holds. That completes the proof of the Corollary.
\end{proof}

\begin{proof}[\textbf{Proof of Corollary 2.2}]
The process $y_t$ can be written by
\begin{equation}\label{ad05}
y_t=\Pi_{s_t}\mathsf{Y}_{t-1}+\xi_t^*,
\end{equation}
where the residual process is given by $\xi_t^*:=\mathsf{G}_t\xi$. The residual process has the following covariance matrix
\begin{equation*}\label{•}
\Sigma_t^*:=\mathsf{G}_t\Sigma_t\mathsf{G}_t'=
\begin{bmatrix}
\Sigma_{11,t} & \Sigma_{12,t}G_t'\\
G_t\Sigma_{21,t} & G_t\Sigma_{22,t}G_t'
\end{bmatrix}.
\end{equation*}
The constraints can be written by
\begin{equation*}\label{•}
\mathbb{\tilde{E}}\big[\exp\{G_t^{-1}\big(\eta_t^*-\hat{\theta}_{2,t}^*\big)\}|\mathcal{H}_{t-1}\big]=i_{n_x},~~~t=1,\dots,T,
\end{equation*}
where $\eta_t^*:=G_t\eta_t$ and $\hat{\theta}_{2,t}^*:=G_t\hat{\theta}_{2,t}$. If we use Corollary 1 for input parameters $R_{2,t}=G_t^{-1}$, $\hat{\theta}_{2,t}=\hat{\theta}_{2,t}^*$, $\Sigma_{12,t}=\Sigma_{12,t}G_t'$, and $\Sigma_{22,t}=G_t\Sigma_{22,t}G_t'$, then we obtain equation \eqref{ad03}. Equations \eqref{01017} and \eqref{ad05} imply equation \eqref{ad04}. That completes the proof of the Corollary.
\end{proof}

\begin{proof}[\textbf{Proof of Theorem 2}]
It is clear that conditional on information $\mathcal{H}_0$ a joint density function of the random vector $y=(y_1',\dots,y_T')'$ is given by
\begin{equation*}\label{01147}
f(y|\mathcal{H}_0)=\frac{1}{(2\pi)^{nT/2}\prod_{t=1}^T|\Sigma_t|^{1/2}}\exp\bigg\{-\frac{1}{2}\sum_{t=1}^T\big(y_t-\Pi_{s_t} \mathsf{Y}_{t-1}\big)'\Sigma_t^{-1}\big(y_t-\Pi_{s_t} \mathsf{Y}_{t-1}\big)\bigg\}
\end{equation*}
Let $S:=\text{vec}\big(y,\Pi_{\hat{s}},\Gamma_{\hat{s}},\hat{s},\mathsf{P}\big)$ be a vector that consists of the random vector $y$, duplication removed random coefficient matrices $\Pi_{\hat{s}}$ and $\Gamma_{\hat{s}}$, random transition matrix $\mathsf{P}$, and duplication removed random regime vector $\hat{s}$. Then a joint density function of the random vector $S$ given $\mathcal{F}_0$ can be represented by
\begin{equation*}\label{01148}
f\big(S |\mathcal{F}_0\big)=f(y |\mathcal{H}_0)\times f\big(S_* |\mathcal{F}_0\big)
\end{equation*}
under the real probability measure $\mathbb{P}$, where $S_*:=\text{vec}(\Pi_{\hat{s}},\Gamma_{\hat{s}},\hat{s},\mathsf{P})$. Thus, conditional on initial information $\mathcal{F}_0$ a joint distribution of the random vector $S$ is given by
\begin{eqnarray*}\label{01149}
&&\mathbb{\tilde{P}}\big[S\in B\big|\mathcal{F}_0\big]=\int_B L_T(y|\mathcal{F}_0)f\big(S |\mathcal{F}_0\big) dS\\
&&= \int_Bc\exp\bigg\{-\frac{1}{2}\sum_{t=1}^T\big(y_t-\Pi_{s_t} \mathsf{Y}_{t-1}-\theta_t\big)'\Sigma_t^{-1}\big(y_t-\Pi_{s_t} \mathsf{Y}_{t-1}-\theta_t\big)\bigg\}\times f\big(S_* |\mathcal{F}_0\big)dS\nonumber
\end{eqnarray*}
under the probability measure $\tilde{\mathbb{P}}$, where $c:=\frac{1}{(2\pi)^{nT/2}\prod_{t=1}^T|\Sigma_t|^{1/2}}$ and $B\in \mathcal{B}\big(\mathbb{R}^{Tn+d_*}\big)$ is an any Borel set with $d_*:=(np+k)nr_{\hat{s}}+(n_*p_*+k_*)n_*r_{\hat{s}}+T+N(N+1)$ is a dimension of the random vector $S_*$. Therefore, one can conclude that conditional on $\mathcal{F}_0$ a joint distribution of the random vector $S_*$ is same for probability measures $\mathbb{\tilde{P}}$ and $\mathbb{P}$, that is, for all Borel set $B_*\in \mathcal{B}\big(\mathbb{R}^{d_*}\big)$, 
\begin{equation}\label{01150}
\mathbb{\tilde{P}}\big[S_*\in B_*\big|\mathcal{F}_0\big]=\mathbb{P}\big[S_*\in B_*\big|\mathcal{F}_0\big].
\end{equation}
Since random vectors $\text{vec}(\Pi_{\hat{s}},\Gamma_{\hat{s}})$ and $\text{vec}(\mathsf{P})$ are independent given $\hat{s}$ and $\mathcal{F}_0$ under the real probability measure $\mathbb{P}$, equation \eqref{01150} can be written by
\begin{equation*}\label{01151}
\tilde{\mathbb{P}}\big[\text{vec}(\Pi_{\hat{s}},\Gamma_{\hat{s}})\in B\big|s,\mathsf{P},\mathcal{F}_0\big]\tilde{\mathbb{P}}\big[\text{vec}(s,\mathsf{P})\in D\big|\mathcal{F}_0\big]=\mathbb{P}\big[\text{vec}(\Pi_{\hat{s}},\Gamma_{\hat{s}})\in B\big|\hat{s},\mathcal{F}_0\big]\mathbb{P}\big[\text{vec}(s,\mathsf{P})\in D\big|\mathcal{F}_0\big],
\end{equation*}
where $B\in\mathcal{B}(\mathbb{R}^{d})$ and $D\in\mathcal{B}(\mathbb{R}^{T+N(N+1)})$ are Borel sets. Therefore, if we take $B=\mathbb{R}^{d}$ in above equation, then conditional on the initial information $\mathcal{F}_0$, a distribution of the random vector $\text{vec}(s,\mathsf{P})$ is same for the probability measures $\mathbb{\tilde{P}}$ and $\mathbb{P}$. Consequently, we have that
\begin{equation*}\label{01152}
\tilde{\mathbb{P}}\big[\text{vec}(\Pi_{\hat{s}},\Gamma_{\hat{s}})\in B\big|s,\mathsf{P},\mathcal{F}_0\big]=\mathbb{P}\big[\text{vec}(\Pi_{\hat{s}},\Gamma_{\hat{s}})\in B\big|\hat{s},\mathcal{F}_0\big].
\end{equation*}
Thus, for given regime--switching vector $s$, transition probability matrix $\mathsf{P}$, and initial information $\mathcal{F}_0$, a distribution of the random vector $\text{vec}(\Pi,\Gamma)$ under the risk--neutral probability measure $\mathbb{\tilde{P}}$ equals for given duplication removed regime--switching vector $\hat{s}$ and initial information $\mathcal{F}_0$, a distribution of the random vector $\text{vec}(\Pi_{\hat{s}},\Gamma_{\hat{s}})$ under the real probability measure $\mathbb{P}$. Moreover, from equation \eqref{01146} we can conclude that conditional on information $\mathcal{H}_0$ a joint distribution of the random vector $y$ is given by
\begin{equation*}\label{01153}
y~|~\mathcal{H}_0 \sim \mathcal{N}\Big(\Psi^{-1}\delta,\Psi^{-1}\Sigma(\Psi^{-1})'\Big)
\end{equation*} 
under the risk--neutral probability measure $\tilde{\mathbb{P}}$. Thus equation \eqref{01019} holds. Thanks to the well--known formula of a partitioned matrix's inverse, we get that
$$\Psi^{-1}=\begin{bmatrix}
\Psi_{11}^{-1} & 0\\
C_{21} & \Psi_{22}^{-1}
\end{bmatrix},$$
where $C_{21}=-\Psi_{22}^{-1}\Psi_{21}\Psi_{11}^{-1}.$ Consequently, we obtain that
$$\Psi^{-1}\delta=\begin{bmatrix}
\Psi_{11}^{-1}\delta_1\\
C_{21}\delta_1+\Psi_{22}^{-1}\delta_2
\end{bmatrix}$$
and 
\begin{eqnarray*}
\Psi^{-1}\Sigma(\Psi^{-1})'=\begin{bmatrix}
\Psi_{11}^{-1}\bar{\Sigma}_t(\Psi_{11}^{-1})' & \Psi_{11}\bar{\Sigma}_t C_{21}'\\
C_{21}\bar{\Sigma}_t(\Psi_{11}^{-1})' & C_{21}\bar{\Sigma}_t C_{21}'+\Psi_{22}^{-1}\bar{\Sigma}_t^c(\Psi_{22}^{-1})'
\end{bmatrix}.
\end{eqnarray*}
So equations \eqref{01020} and \eqref{01021} holds. From the well--known formula of the conditional distribution of multivariate random vector, one can obtain equations \eqref{01022} and \eqref{01023}. The proof is complete.
\end{proof}

\begin{proof}[\textbf{Proof of Lemma 1}]
Using the fact that $(X_1,Y_1),\dots,(X_n,Y_n)$ are identically distributed random vectors and iterated expectation formula, namely, $\mathbb{E}\big[\mathbb{E}[h(X,Y)|Y]\big]=\mathbb{E}[h(X,Y)]$ we obtain 
$$\mathbb{E}[\tau_1]=\mathbb{E}[\tau_2]=\mathbb{E}[h(X,Y)].$$
On the other hand, as $(X_1,Y_1),\dots,(X_n,Y_n)$ are independent copy of $(X,Y)$, we have
$$\mathrm{Var}(\tau_1)=\frac{1}{n}\mathrm{Var}[h(X,Y)]=\frac{1}{n}\Big(\mathbb{E}[h(X,Y)^2]-\mathbb{E}^2[h(X,Y)]\Big)$$
and
$$\mathrm{Var}(\tau_2)=\frac{1}{n}\mathrm{Var}[g(Y)]=\frac{1}{n}\Big(\mathbb{E}[g(Y)^2]-\mathbb{E}^2[h(X,Y)]\Big).$$
By applying the Jensen's inequality for a function $\varphi(x)=x^2$, one obtains 
$$\mathbb{E}[g(Y)^2]=\mathbb{E}\big[\mathbb{E}^2[h(X,Y)|Y]\big]\leq \mathbb{E}\big[\mathbb{E}[h(X,Y)^2|Y]\big]=\mathbb{E}[h(X,Y)^2].$$
Thus, the inequality $\mathrm{Var}(\tau_1)\geq \mathrm{Var}(\tau_2)$ holds.
\end{proof}

\begin{proof}[\textbf{Proof of Lemma 2}]
Firstly, let us consider the expectation $\mathbb{E}\big[(X-K)^+\big]$. Since $X\sim \mathcal{N}(\mu,\sigma^2)$, it is obvious that
\begin{equation*}\label{01154}
\mathbb{E}\big[(X-K)^+\big]=\int_K^\infty \frac{x-K}{\sqrt{2\pi\sigma^2}}\exp\bigg\{-\frac{(x-\mu)^2}{2\sigma^2}\bigg\}dx.
\end{equation*}
Let us introduce $z:=(x-\mu)/\sigma$ and $\bar{K}:=(K-\mu)/\sigma$. Then, we have
\begin{equation*}\label{01155}
\mathbb{E}\big[(X-K)^+\big]=\sigma\int_{\bar{K}}^\infty \frac{ z}{\sqrt{2\pi}}\exp\bigg\{-\frac{z^2}{2}\bigg\}dz+(\mu-K)\int_{\bar{K}}^\infty \frac{1}{\sqrt{2\pi}}\exp\bigg\{-\frac{z^2}{2}\bigg\}dz.
\end{equation*}
Since the standard normal density function is symmetric at the origin, the second integral of the above equation equals $(\mu-K)\Phi(-\bar{K})$. Thus, equation \eqref{01035} holds. Similarly, one can obtain \eqref{01036}.
\end{proof}

\begin{proof}[\textbf{Proof of Lemma \ref{lem04}}]
By the conditional probability formula, one gets that
\begin{equation*}\label{01162}
\tilde{f}(\bar{y}_t,\Pi_{\hat{s}},\Gamma_{\hat{s}},s,\mathsf{P}|\mathcal{F}_0)=\tilde{f}(\bar{y}_t|\Pi_{\hat{s}},\Gamma_{\hat{s}},s,\mathsf{P},\mathcal{F}_0)\tilde{f}(\Pi_{\hat{s}},\Gamma_{\hat{s}}|s,\mathsf{P},\mathcal{F}_0)\tilde{f}(s,\mathsf{P}|\mathcal{F}_0).
\end{equation*}
Due to Theorem 2, the first, second, and third terms of the right--hand side of the above equation equal $\tilde{f}(\bar{y}_t|\Pi_{\alpha},\Gamma_{\alpha},\bar{s}_t,\mathcal{F}_0)$, $f(\Pi_{\hat{s}},\Gamma_{\hat{s}}|\hat{s},\mathcal{F}_0)$, and $f(s,\mathsf{P}|\mathcal{F}_0)$, respectively. Consequently, by equation \eqref{08011}, we have that
\begin{equation}\label{01163}
\tilde{f}(\bar{y}_t,\Pi_{\hat{s}},\Gamma_{\hat{s}},s,\mathsf{P}|\mathcal{F}_0)=\tilde{f}(\bar{y}_t|\Pi_{\alpha},\Gamma_{\alpha},\bar{s}_t,\mathcal{F}_0)f(\Pi_{\alpha},\Gamma_{\alpha}|\alpha,\mathcal{F}_0)f_*(\Pi_{\delta},\Gamma_{\delta}|\delta,\mathcal{F}_0)f(s,\mathsf{P}|\mathcal{F}_0).
\end{equation}
If we take integral from the above equation with respect to $(\Pi_{\hat{s}},\Gamma_{\hat{s}},s,\mathsf{P})$, then we find conditional density of the random vector $\bar{y}_t$ under the risk--neutral probability $\mathbb{\tilde{P}}$, namely,
\begin{eqnarray}\label{01166}
\tilde{f}(\bar{y}_t|\mathcal{F}_0)=\sum_{\bar{s}_t}\bigg(\int_{\Pi_{\alpha},\Gamma_{\alpha}}\tilde{f}(\bar{y}_t|\Pi_{\alpha},\Gamma_{\alpha},\bar{s}_t,\mathcal{F}_0)f(\Pi_{\alpha},\Gamma_{\alpha}|\alpha,\mathcal{F}_0)d\Pi_\alpha d\Gamma_\alpha\bigg)f(\bar{s}_t|\mathcal{F}_0).
\end{eqnarray}
Dividing equation \eqref{01163} by equation \eqref{01166}, one obtains equation \eqref{07042}. If we integrate equation \eqref{07042} by $\mathsf{P}$, then we get equation \eqref{ad001}.
\end{proof}

\begin{proof}[\textbf{Proof of Lemma 4}]
Firstly, let $\alpha, \beta\in\mathbb{R}$ be constants, $X$ be a normal random variable with parameters $\mu$ and $\sigma^2$, that is, $X\sim \mathcal{N}(\mu,\sigma^2)$, and $Z$ be a standard normal random variable, which is independent of the random variable $X$. Then, we have
$$\mathbb{E}\big[\Phi(\alpha X+\beta)\big]=\mathbb{E}\big[\mathbb{P}[Z\leq \alpha X+\beta|X]\big]=\mathbb{P}[Z-\alpha X\leq \beta].$$
Because the random variable $Z-\alpha X$ follows normal distribution with mean $-\alpha\mu$ and variance $1+\alpha^2\sigma^2$, we obtain that 
\begin{equation}\label{eq092}
\mathbb{E}\big[\Phi(\alpha_1X+\beta_1)\big]=\Phi\bigg(\frac{\alpha_1\mu+\beta_1}{\sqrt{1+\alpha_1^2\sigma^2}}\bigg).
\end{equation}
Secondly, let us consider an expectation, which will be used to prove the Lemma $\mathbb{E}\big[\exp\big\{\alpha_1X+\beta_1\big\}\Phi(\alpha_2X+\beta_2)\big]$ for constants $\alpha_1,\alpha_2,\beta_1,\beta_2\in \mathbb{R}$.
For exponents of the expectation, observe that
\begin{equation}\label{eq093}
\alpha_1x+\beta_1-\frac{(x-\mu)^2}{2\sigma^2}=\alpha_1\mu+\beta_1+\frac{\alpha_1^2\sigma^2}{2}-\frac{(x-\mu-\alpha_1\sigma^2)^2}{2\sigma^2}.
\end{equation}
The last term corresponds to normal random variable with parameters $\mu+\alpha_1\sigma^2$ and $\sigma^2$. As a result, we get that
$$\mathbb{E}\Big[\exp\big\{\alpha_1X+\beta_1\big\}\Phi(\alpha_2X+\beta_2)\Big]=\exp\bigg\{\alpha_1\mu+\beta_1+\frac{\alpha_1^2\sigma^2}{2}\bigg\}\mathbb{E}\big[\Phi(\alpha_2Y+\beta_2)\big],$$
where $Y\sim \mathcal{N}(\mu+\alpha_1\sigma^2,\sigma^2)$. By using equation \eqref{eq092}, we get that
\begin{equation}\label{eq094}
\mathbb{E}\Big[\exp\big\{\alpha_1X+\beta_1\big\}\Phi(\alpha_2X+\beta_2)\Big]=\exp\bigg\{\alpha_1\mu+\beta_1+\frac{\alpha_1^2\sigma^2}{2}\bigg\}\Phi\bigg(\frac{\alpha_2\mu+\beta_2+\alpha_1\alpha_2\sigma^2}{\sqrt{1+\alpha_2^2\sigma^2}}\bigg),
\end{equation}
Thirdly, the iterated expectation formula implies that
\begin{equation}\label{eq095}
\mathbb{E}\Big[\big(e^{X_1}-e^{X_2}\big)^+\Big]=\mathbb{E}\Big[\mathbb{E}\big[\big(e^{X_1}-e^{X_2}\big)^+\big|X_1\big]\Big].
\end{equation}
In the first step, let us consider the conditional expectation of equation \eqref{eq095}. According to the well--known conditional distribution formula of the multivariate normal distribution, we have 
$$X_2~|~X_1\sim \mathcal{N}\big(\mu_{2.1}(X_1),\sigma_{22.1}^2\big),$$
where $\mu_{2.1}(X_1):=\mu_2+\sigma_{12}/\sigma_1^2(X_1-\mu_1)$ is a conditional mean and $\sigma_{22.1}^2:=\sigma_2^2-\sigma_{12}^2/\sigma_1^2$ is a conditional variance of the random variable $X_2$ given $X_1$. Let us denote a density function of the random variable $X_2$ given $X_1$ by $\phi(x_2|X_1)$. Then, one get that
\begin{equation}\label{eq096}
\mathbb{E}\big[\big(e^{X_1}-e^{X_2}\big)^+\big|X_1\big]=e^{X_1}\int_{-\infty}^{X_1}\phi(x_2|X_1)dx_2-\int_{-\infty}^{X_1}e^{x_2}\phi(x_2|X_1)dx_2.
\end{equation}
For the first term of the right--hand side of the above equation, we have
\begin{equation}\label{eq097}
e^{X_1}\int_{-\infty}^{X_1}\phi(x_2|X_1)dx_2=e^{X_1}\Phi\bigg(\frac{X_1-\mu_{2.1}(X_1)}{\sqrt{\sigma_{22.1}}}\bigg).
\end{equation}
For the second term of the right--hand side of equation \eqref{eq096}, by the completing the square method, see equation \eqref{eq093}, one has
\begin{equation}\label{eq098}
\int_{-\infty}^{X_1}e^{x_2}\phi(x_2|X_1)dx_2=\exp\bigg\{\mu_{2.1}(X_1)+\frac{\sigma_{22.1}}{2}\bigg\}\Phi\bigg(\frac{X_1-\mu_{2.1}(X_1)-\sigma_{22.1}}{\sqrt{\sigma_{22.1}}}\bigg).
\end{equation}
Finally, since $\mu_{2.1}(X_1)$ is the linear function for its argument $X_1$, using equation \eqref{eq094} for equations \eqref{eq097} and \eqref{eq098}, one completes the proof of the Lemma. 
\end{proof}

\begin{proof}[\textbf{Proof of Lemma 5}]
By taking $\mu_1=\mu$, $\mu_2=\ln(K)$, $\sigma_1^2=\sigma^2$, and $\sigma_{12}=\sigma_2^2=0$ in Lemma \ref{lem05}, one obtains the first expectation, given by equation \eqref{01109}. On the other hand, by taking $\mu_1=\ln(K)$, $\mu_2=\mu$, $\sigma_2^2=\sigma^2$, and $\sigma_{12}=\sigma_1^2=0$ in Lemma \ref{lem05}, we find the second expectation, given by equation \eqref{01110}.
\end{proof}

\bibliographystyle{apacite}
\bibliography{References}

\end{document}